**Automated deep reinforcement learning for real-time scheduling strategy of multi-energy system integrated with post-carbon and direct-air carbon captured system.**

Tobi Michael Alabi [a, b, c], Nathan P. Lawrence [c, d], Lin Lu [b, *], Zaiyue Yang [a, *], R. Bhushan Gopaluni [c, *]

[a] *Department of Mechanical and Energy Engineering, Southern University of Science and Technology, Shenzhen, China (tobi.alabi@connect.polyu.hk)*

[b] *Renewable Energy Research Group (RERG) Department of Building Environment and Energy Engineering, The Hong Kong Polytechnic University, Hong Kong, China,*

[c] *Data Analytics and Intelligent System (DAIS) Laboratory, Department of Chemical and Biological Engineering, University of British Columbia, Vancouver, BC, Canada*

[d] *Department of Mathematics, University of British Columbia, Vancouver, BC, Canada*

**Corresponding Authors: Zaiyue Yang, yangzy3@sustech.edu.cn; Lin Lu, email: vivien.lu@polyu.edu.hk; R.B. Gopaluni, email: bhushan.gopaluni@ubc.ca*

## Abstract

The carbon-capturing process with the aid of $CO_2$ removal technology (CDRT) has been recognised as an alternative and a prominent approach to deep decarbonisation. However, the main hindrance is the enormous energy demand and the economic implication of CDRT if not effectively managed. Hence, a novel deep reinforcement learning agent (DRL), integrated with an automated hyperparameter selection feature, is proposed in this study for the real-time scheduling of a multi-energy system (MES) coupled with CDRT. Post-carbon capture systems (PCCS) and direct-air capture systems (DACS) are considered CDRT. Various possible configurations are evaluated using real-time multi-energy data of a district in Arizona, the United States, and CDRT parameters from manufacturers' catalogues and pilot project documentation. The simulation results validate that an optimised soft-actor critic (SAC) DRL algorithm outperformed the Twin-delayed deep deterministic policy gradient (TD3) algorithm due to its maximum entropy feature. We then trained four (4) SAC DRL agents, equivalent to the number of considered case studies, using optimised hyperparameter values and deployed them in real time for evaluation. The results show that the proposed DRL agent can meet the prosumers' multi-energy demand and schedule the CDRT energy demand economically without specified constraints violation. Also, the proposed DRL agent outperformed rule-based scheduling by 23.65%. However, the configuration with PCCS and solid-sorbent DACS is considered the most suitable configuration with a high $CO_2$ captured-released ratio (CCRR) of 38.54, low $CO_2$ released indicator (CRI) value of 2.53, and a 36.5% reduction in CDR cost due to waste heat utilisation and high absorption capacity of the selected sorbent. However, the adoption of CDRT is not economically viable at the current carbon price. Finally, we showed that CDRT would be attractive at a carbon price of 400-450USD/ton with the provision of tax incentives by the policymakers.

## 1.0 Introduction

### *1.1 Motivation and Background*

Carbon neutrality has become an ambitious goal; achieving the milestone on or before 2060 is of utmost importance [1]. The passion for its feasibility is evident with the rapid transformation of the energy sector. Specifically, through huge investments in renewable energy, electrification of thermal production, decarbonisation of the transportation sector, and the synergy of various energy components that is termed multi-energy systems (MES) [2, 3]. Further, a zero-carbon multi-energy system (ZCMES) that strictly aims at achieving negative emissions with MES has been proposed by curbing the associated $CO_2$ within the framework or as a result of dependency on the external grid [2, 4].

Meanwhile, despite the commissioning of 261GW of renewable energy plants in 2020, the reality on the ground shows that fossil fuel combustion (coal, oil, natural gas, and other fuel) still accounts for 61.3% of the world's electricity generation in 2020 while renewable energy penetration is 27.7% [5]. The reasons for these are 1) renewable energy installations are restricted by available space, and its large installation has received several criticisms from environmentalists under the prospect of desert encroachment and natural habitat destabilisation [6]; 2) renewable energy real-time performance is greatly influenced by weather behaviour, and total reliance on its prediction can be disastrous; a typical example was the shutting down of wind turbine farms in Texas due to extreme weather conditions [7], and 3) the fear of national economy and international trade relationship collapse if fossil fuel production is overhaul since fossil is the primary international products of some notable countries [8]. Besides, some of the innovative zero-emission and energy storage technologies are still expensive due to the scarce materials required for their functionality [9, 10]. Therefore, while renewable energy penetration is increasing every year, as recorded in 2020, the exploration of achieving zero-emission with the existing technologies by harnessing their associated $CO_2$ emission via the carbon capture approach is worthy of research investment. Furthermore, the optimal economic dispatch of ZCMES while ensuring its feasibility and reliability during operation is another crucial area that cannot be disregarded. Hence, an optimal control platform that is accurate, economical, has uncertainty consideration and facilitates real-time deployment is required.

### *1.2 Related research works*

The feasibility of zero-carbon in the energy sector can be achieved through 1) the use of zero-emission technologies, 2) the integration of carbon capture technologies, and 3) the optimal configuration of zero-emission and carbon capture technologies. Numerous researchers on zero-emission energy units have invested invaluable efforts in their feasibility and efficiency[11]. Meanwhile, their economic implication, such as huge capital and maintenance costs, are the primary disadvantage [2]. The carbon capture approach is another carbon mitigation approach that has received attention. The technique is divided into the pre-combustion process, post-carbon capture (PCC), direct air capture (DAC), and oxyfuel combustion [12]. However, PCC and DAC are the most adopted methods and have been tested on a pilot scale and commercialised [13].

PCC is a carbon-dioxide removal (CDR) technology that removes the $CO_2$ in the exhaust gas of carbon emission equipment such as coal-fired power plants and natural gas-fired equipment [14]. Specifically, the flue gas passes through a connection pipe, as illustrated in Fig. 1, and contacts the solvent (aqueous sorbent or solid), usually an amine solution inside the PCCS. This chemical solution can absorb up to 85% of the flue gas $CO_2$ content while allowing the remaining gaseous component (usually oxygen and nitrogen) to be released freely into the atmosphere. The $CO_2$-rich amine solution is regenerated in the desorber by stripping the $CO_2$ out of the liquid using thermal energy [15], and then the lean amine solution is recycled in the absorber while the stripped $CO_2$ in liquid form is compressed, cooled, and stored or transported. Some notable studies have incorporated this technology into the MES framework and evaluated its influence on the system's performance. Zhang et al. [16] evaluated PCC integration's efficiency and economic implication with a liquified natural gas plant, Solid oxide fuel cell, and compressed air storage. The obtained results show that nearly 100% carbon capture of the liquified natural gas plant is possible with 1.69$/h reduction in total cost. The optimal dispatch of MES was examined by G. Zhang et al. [17] by utilising the carbon capture by PCC for sync gas production via power-to-gas (P2G) technology and power generation by supercritical $CO_2$ (S-$CO_2$) cycle; the carbon emission of the system was reduced by 89.5% with 1% increase of the energy efficiency. In Berge et al. [18], PCC was coupled with P2G to achieve a low-carbon

economy of MES. X. Zhang et al., [19] also integrate PCC into MES infrastructure with the aim of utilizing the captured CO2 as synthetic fuel. G. Zhang et al. [20] went a step further with the introduction of supercritical-CO2 unit to aid power generation using the captured CO2 within the MES. However, none of these studies considered the PCC capacity constraint on the amount of captured CO2 and the possible release CO2 to the environment. Moreover, the proposed energy management is not autonomous which be difficult to execute in real-time.

Similarly, Berger et al. [21] evaluated the role of CDRT and P2G as a decarbonisation strategy; they asserted that the CDRT plays an enabling role, but their high energy penalty effects may exacerbate dependency on fossil fuels. Besides, PCC cannot guarantee a complete harness of $CO_2$ since its efficiency is not 100%; the escaped $CO_2$ and the surrounding $CO_2$ are excluded, and the installed capacity restricts the quantity of absorbed CO2. Another germane omission in the studies mentioned above is the optimal power scheduling of each power generation unit for PCC consumption to optimise its huge energy penalty. Moreover, few studies have integrated PCC as part of MES components.

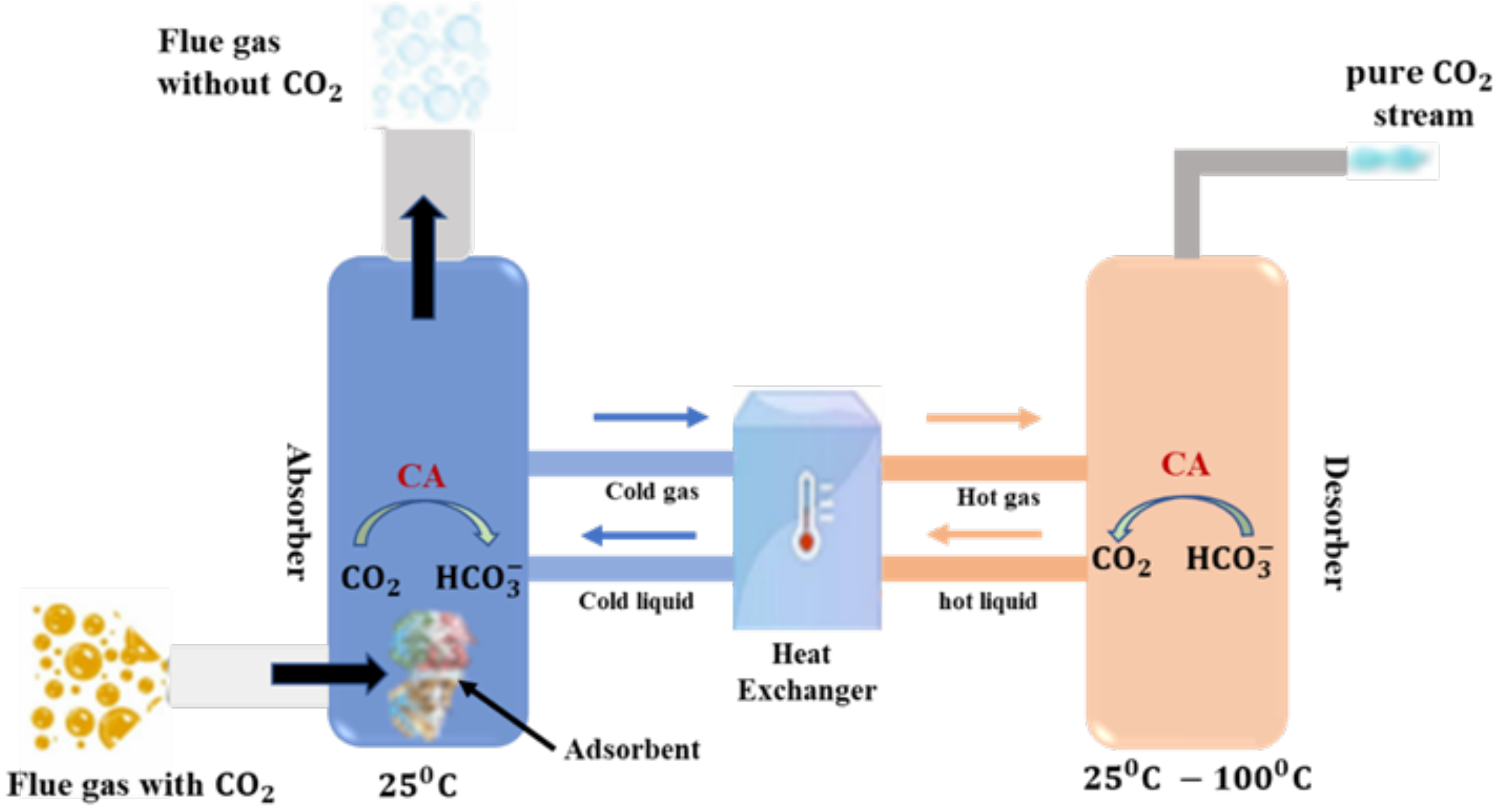

**Figure 1**. Post-carbon capture (PCC) system.

DAC is another mature CDR technology that aids the feasibility of negative emissions in the energy sector. Compared to PCC, the primary aim is to remove $CO_2$ in the atmosphere through suction of the surrounding air by ventilators and $CO_2$ collectors, as illustrated in Fig 2, binding the $CO_2$ present in the air through the absorption process when they are in contact with the sorbent (aqueous solution or solid sorbent) in the vacuum system. This is followed by stripping $CO_2$ from the lean solution in the regenerator/desorber using thermal energy and sorbent recycling based on a cyclic process [22]. Tom et al. [23] evaluated various DAC systems' life cycle assessment analysis. They showed that by integrating DAC with high solar radiation and high usage of waste heat for $CO_2$ regeneration in the desorber, the $CO_2$ removal efficiency could be up to 79-91%. The main types of DAC are absorption and adsorption DAC: The former uses aqueous solvent solutions such as alkali materials to remove $CO_2$, while the latter uses solid sorbent [24]. A large pilot study of absorption DAC has been established in various North American markets by Carbon Engineering (Canada) [25], while Climeworks has been the main leader in solid-sorbent DAC in Europe [26].

A typical example is the pilot projects in Hinwil, Switzerland and Hellisheidi in Iceland that achieved negative $CO_2$ emissions in 2017 [27]. Nonetheless, the main drawback of DAC is the huge energy demand for the regeneration process and the absorption capacity of the selected sorbent. For instance, an aqueous solvent DAC requires about $900^0$C temperature for the regeneration process accompanied by high water loss. At the same time, a solid sorbent DAC (also termed low-temperature DAC) needs around $100^0$C for the $CO_2$ splitting that can be provided by low-grade heat, e.g. waste heat [28]. However, the aqueous solvent has a low absorption capacity compared to solid sorbent, which contributes to high solvent consumption at each time slot [29]. A comparative life cycle analysis of aqueous and solid sorbents DAC was carried out in [24, 29]. Meanwhile, despite the possibility of DAC complementing PCC to achieve deep decarbonisation, the optimal configuration has rarely been considered. In fact, to the best of the authors' knowledge, our study is the first to integrate these two CDR technologies as part of ZCMES.

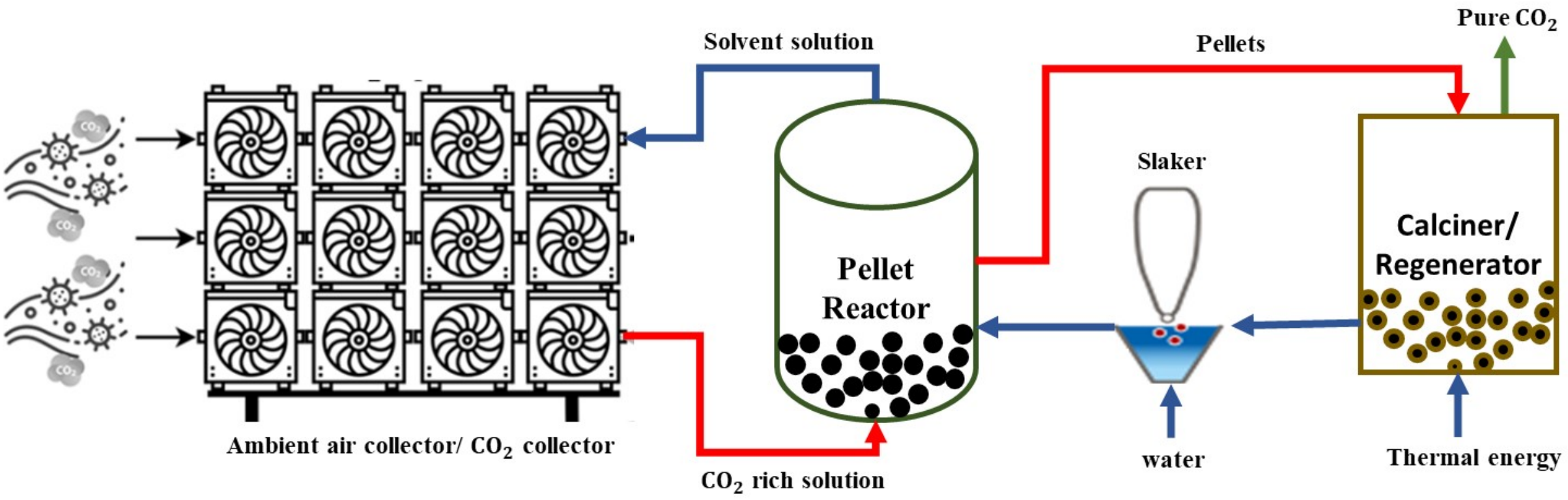

**Figure 2.** Direct-air carbon capture system.

Conventionally, the optimal scheduling strategy has been implemented using mathematical programming in extant studies [1, 30]. In contrast, the conventional control strategies suffer from two major setbacks, which are 1) the inability to be deployed in real-time; and 2) local optimum results due to a reduction in model complexity that sacrifices accuracy for computational cost. Reinforcement learning (RL) tackles the inadequacy of the control mentioned above techniques and obtain results that is close to global optimum [31]. Firstly, it utilises the Bellman Equation to achieve global optimum, which entails the combination of the current state's immediate reward with the next state reward and the cumulative reward from the next step to the terminal state. Secondly, it eliminates the reduced-order model dependency with a model-free approach. Combining the mentioned features enables less online computation time, optimum global possibility, and real-time deployment. The central intuition of RL is the selection of optimal policy that maximises the cumulative reward; this is done by an iterative interaction between the RL agent and the environment, as shown in Fig 3, while the agent parameters are updated during the learning process. In this study, neural networks (NN) is use in the deep RL algorithm to address the curse of dimensionality (more details in Section 3.2.1) Remarkably, some studies have applied DRL to the optimal scheduling of energy systems MES, as reviewed in [31, 32]. However, throughout our research database consultations, no study has applied DRL for the optimal control of carbon capture technology.

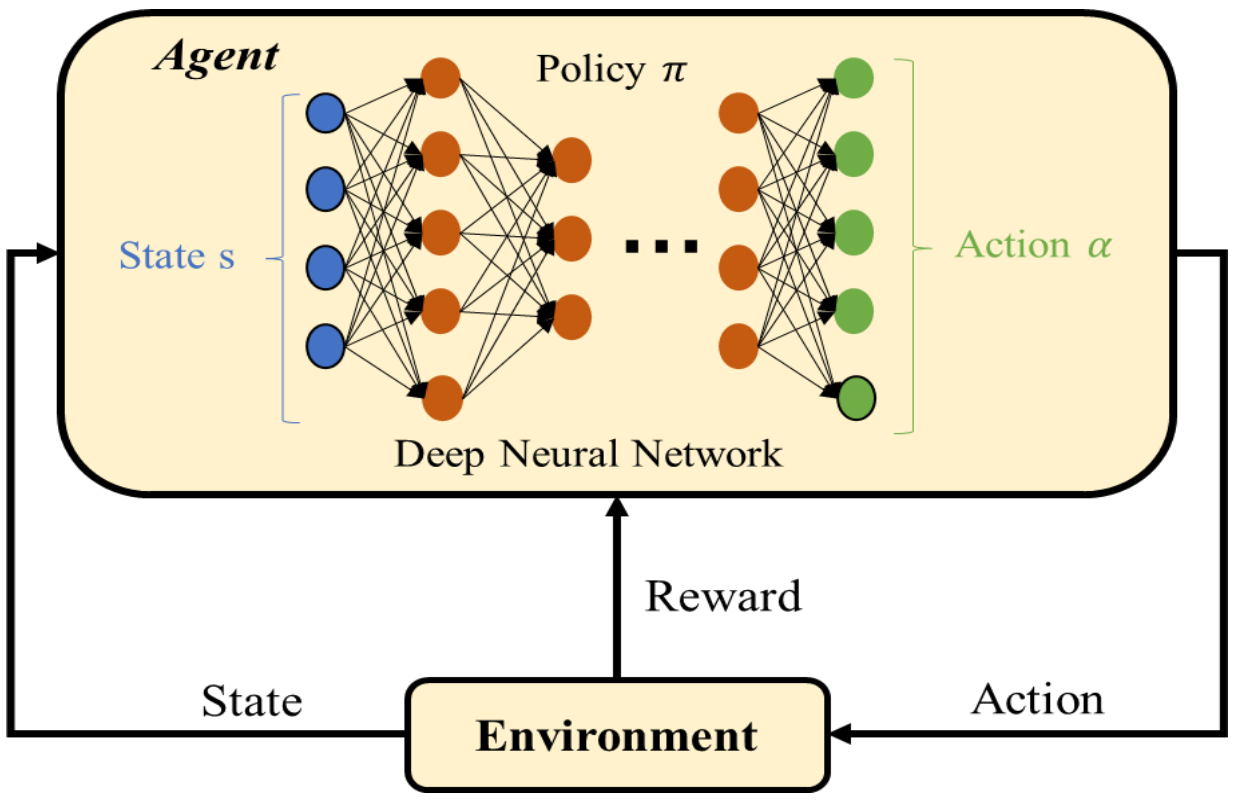

**Figure 3**. Schematic illustration of RL interaction with the environment.

*1.3 Novelty Contributions*

Compared to the studies mentioned above, this paper is the first to consider the integration of PCC and DAC as a promising strategy for the feasibility of ZCMES. Moreover, a DRL algorithm that can be deployed in real-time to achieve an optimal prediction control strategy is proposed. The main contributions and technical advancement of the study are summarised below:

1. PCC and DAC, the leading CDR technologies, are considered part of the energy system for the feasibility of ZCMES.

2. A comparison between the techno-economic performance of aqueous solvent DAC and solid sorbent DAC on the overall performance of ZCMES is evaluated.
3. A DRL algorithm with an automatic hyperparameter tuning feature is developed for the real-time scheduling of the proposed ZCMES.

## 2.0 System description and mathematical model

A system configuration that has the potential of achieving zero-emission is proposed in this study, as shown in Fig 4. The full utilisation of available renewable resources (wind power WT and solar generation PV) is a priority. However, the highly stochastic nature of the resources and available space constraints are the primary limitation. Besides, for large multi-energy consumers, installing WT and PV as the primary energy source may not be an excellent economic approach. Hence, a gas turbine (GT) and a coal-fired plant (CFP) that function as a cogeneration plant are introduced. GT and CFP generate electricity and waste heat simultaneously, with the aim of waste heat recovery and its utilisation to meet part of the thermal demand of the prosumers. Furthermore, two (2) common energy storage methods (battery electrical storage (BES) and thermal energy storage (TES)) are introduced as backup systems. The heat demand of the prosumers is to be provided by a gas boiler (GB) and water-source heat pump (WSHP), while the cooling demand will be met by an electric chiller (EC) and absorption chiller (AC). Remarkably, to achieve carbon neutrality, PCC and DAC system are incorporated.

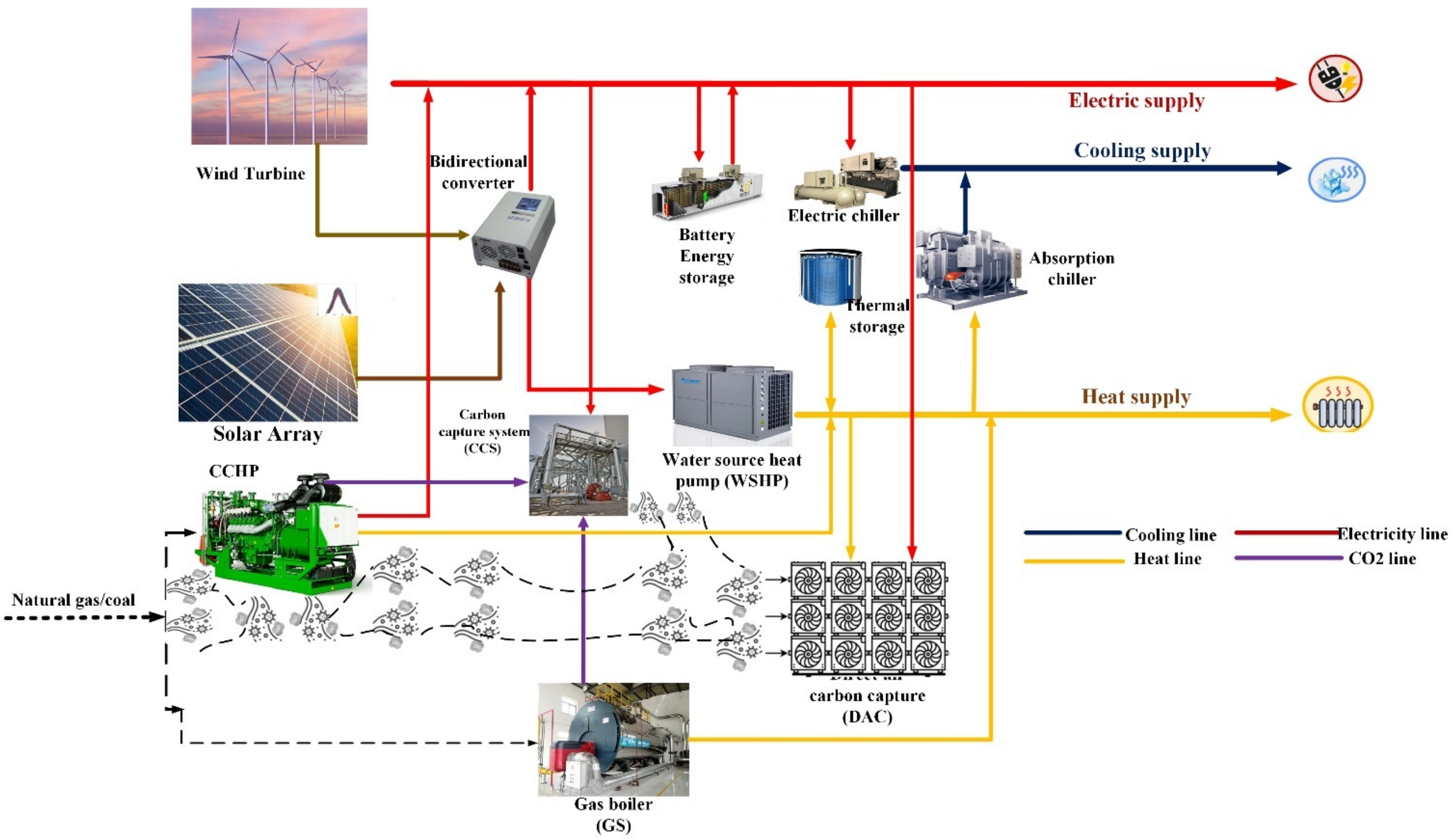

**Figure 4.** The proposed zero-carbon multi-energy system (ZCMES) with a post-carbon and direct-air carbon capture system.

*2.1 Part Load Rate modelling*

Most existing studies on MES modelled the power generation and thermal equipment using a constant efficiency or coefficient of performance (COP) when estimating the power output. However, the status of this equipment deviates from the on-design condition, which is influenced by the load variation at the demand side and some externally influenced factors such as ambient temperature and power output cycling between time steps. Indeed, Huang et al. [33] analysed and compared the MES off-design and on-design operation models. Their validation shows that the scheduling strategy improves with increased operation costs. However, the least operation cost should not be sacrificed for feasibility. Hence, as illustrated in Eq (1) – eq (5), a part load rate (PLR) model is incorporated while modelling GT, CFP, GB, WSHP, EC, and AC to ensure the credibility of the model following the proposed mathematical model in [33, 34]. A detailed model of these units is presented in the Appendix section

$$P_o = P^{r,i}\delta_{i,(t)} \quad (1)$$

$$u_{i(t)}\delta_{i,min} \le \delta_{i,(t)} \le u_{i(t)}\delta_{i,max} \quad (2)$$

$$P_{in,(t)} = {P_o}/{\eta_{i,(t)}} \quad (3)$$

$$\eta_{i,(t)} = \begin{cases} \eta_r \sum_{o \epsilon O} k_o \delta_{i,(t)}, & with\ PLR \\ \eta_r, & without\ PLR \end{cases} \quad (4)$$

Eq (1) denotes the power output of the unit at each time step where $P^{r,i}$ is the rated capacity of device $i$. $P_{in,(t)}$ represents energy input into the device. $\delta_{(t)}$ is the load rate at each time slot with a value from minimum to the maximum loading rate which varies from 0 to 100%, and $u_{i(t)}$ is a binary variable that controls the status of the device where 0 indicates OFF and 1 denotes ON in eq (2). PLR is a non-linear function as shown in eq (3), and $\eta_{(t)}$ is the PLR polynomial expression that is computed in eq (4) where $k_o$ is the vectorized fitting coefficients with respect to loading rate obtained from [33]. When PLR is not considered, $\eta_{(t)}$ will be treated as $\eta_r$ which denotes the rated efficiency/COP.

*2.2 Energy storage system model*

An improved mathematical model of BES and TES, considering its ease of incorporation into the DRL problem is proposed in this study. Eq. 5 describes the estimated charge or discharge power of each energy storage system where $\delta_{x(t)}$ denotes the loading rate at each time step and $P_x^r$ is the equipment capacity rating. The loading rate range is described in Eq. 6, ranging from -1 to +1, where the negative value indicates discharge and the positive value denotes the charging process, the state of charge (SOC) of the system at each time step is computed in Eq. 7. Further, the charging efficiency ($\eta_x$) of the system for either the charging or discharging process is described in Eq. 8. The storage SOC limit at each time step is described in eq. 9. However, the constraint is transformed as illustrated in Eq. 10 using a penalty cost ($\psi_{x,t}$) where $\psi_1$ and $\psi_2$ are the penalty coefficients to ensure adaptation into MDP formulation. Finally, Eq. 11 denotes the charging and discharging power distribution of each energy storage system; during the charging process of BES, the power is supplied by RES ($P_{RES\rightarrow BES,t}$) and GT($P^e_{CHP\rightarrow BES,t}$) while the discharging power is distributed to the electric load ($P_{BES\rightarrow EL,t}$), WSHP ($P_{BES\rightarrow WSHP,t}$), PCC ($P_{BES\rightarrow PCCS,t}$) and DAC ($P_{BES\rightarrow DAC,t}$). A similar approach is also applied to TES, where the charging power is supplied by EB ($P_{EB\rightarrow TES,t}$), GB ($P_{GB\rightarrow TES,t}$), and GT ($P^h_{CHP\rightarrow TES,t}$) and the discharging power is distributed to the heat load ($P_{TES\rightarrow QL,t}$), AC ($P_{TES\rightarrow AC,t}$), and DAC ($P_{TES\rightarrow DAC,t}$). It is worth mentioning that the primary aim of the scheduling problem is to achieve optimal power distribution.

$$P_{x(t)} = \delta_{x(t)} P_x^r, \quad x \in \{BES, TES\}CC \tag{5}$$

$$\delta_{x,min} \leq \delta_{x(t)} \leq \delta_{x,max} \tag{6}$$

$$SoC_{x,t} = SoC_{x,t-1} - \frac{\eta_x P_{x(t)} \Delta t}{Q_x} \tag{7}$$

$$\eta_x = \begin{cases} \eta_{x,ch}, & P_{x,(t)} < 0 \\ 1/\eta_{x,dch}, & P_{x,(t)} \geq 0 \end{cases} \tag{8}$$

$$SOC_{x,min} \leq SOC_{x,t} \leq SOC_{x,max} \tag{9}$$

$$\psi_{x,t} = \begin{cases} \psi_1 e^{(SoC_{min} - SoC_t)}, & SoC_{x,t} < SoC_{x,min} \\ 0, & SOC_{x,min} \leq SOC_{x,t} \leq SOC_{x,max} \\ \psi_2 e^{(SoC_{x,t} - SoC_{x,max})}, & SOC_{x,t} > SOC_{x,max} \end{cases} \tag{5}$$

$$|P_{x\leftarrow(t)}| = \begin{cases} P_{RES\rightarrow BES,t} + P^e_{CHP\rightarrow BES,t}, & P_{BES(t)} < 0 \\ P_{BES\rightarrow EL,t} + P_{BES\rightarrow WSHP,t} + P_{BES\rightarrow PCCS,t} + P_{BES\rightarrow DAC,t}, & P_{BES(t)} \geq 0 \\ P_{EB\rightarrow TES,t} + P_{GB\rightarrow TES,t} + P^h_{CHP\rightarrow TES,t}, & P_{TES,(t)} < 0 \\ P_{TES\rightarrow QL,t} + P_{TES\rightarrow AC,t} + P_{TES\rightarrow DAC,t}, & P_{TES,(t)} \geq 0 \end{cases} \tag{6}$$

*2.3 Post-carbon capture system*

In this study, the flue gas emitted by the emission technologies (GT, CFP, and GB) is harnessed by fitting the exhaust chambers of the system to PCCS. The governing mathematical equations that are applicable in this work are described below. Eq. 12 expresses the total emissions emitted by MES. We assume the environmental $CO_2$ is equivalent to 10% of the MES emission, and the final total emission is computed in Eq. 13. The quantity of $CO_2$ that PCC can capture is estimated in Eq. 14 where $\beta_{pcc}$ is the capturing efficiency, $\delta_{PCC(t)}$ is the loading rate that is constrained by Eq. 15, and $P^r_{PCC}$ is the installed capacity in tons. The electrical consumption of the PCC ($P^{PCC}_{e,t}$) is computed in Eq. 16 where $\delta^p$ is the electricity demand per MWh/ton. Finally, the solvent cost is computed in Eq.17 where $\Psi^{PCC}_{AB}$ is the solvent consumption in kg/$CO_2$ton.

$$QCO^{MES}_{2,t} = GT_{CO2(t)} + D^{CfP}_{CO2(t)} + GB_{CO2(t)} \tag{12}$$

$$QCO^{total}_{2,t} = QCO^{MES}_{2,t} + 0.1QCO^{MES}_{2,t} \tag{13}$$

$$Q^{pcc}_{co2(t)} = \beta_{pcc} \delta_{PCC(t)} P^r_{PCC} \tag{14}$$

$$u_{PCC(t)} \delta_{PCC,min} \leq \delta_{PCC(t)} \leq u_{PCC(t)} \delta_{PCC,max}; \; u_{PCC(t)} \in \{0,1\} \tag{15}$$

$$P^{PCC}_{e,t} = \delta^p Q^{pcc}_{co2(t)} \tag{16}$$

$$\Psi_{PCC,t}^{Absorp} = \Psi_{AB}^{PCC} Q_{co2(t)}^{cap} \quad (17)$$

*2.4 Direct-Air carbon capture system*

Direct-air carbon capture is one of the carbon removal technologies that aid the feasibility of negative emissions. In this study, DAC is introduced to complement the capturing potentials of PCC without the aim of replacing it. The reason for this is not farfetched: 1) The amount of $CO_2$ that PCC can capture is constrained by its installed capacity, which also has economic implications; 2) PCC is not an ideal system that will guarantee 100% efficiency which will lead to the release of some $CO_2$ in the atmosphere. Besides, DAC also captured some of the environmental emissions outside the ZCMES boundary, such as combustion vehicle emissions, and the optimal scheduling of both DAC and PCC will yield an economic benefit. The governing equation of DAC is described in Eq. 18 - Eq. 22. Following a similar capturing process as described for PCC in section 2.3, the only addition here is the computation of the released $CO_2$ that the CDRT does not capture.

$$Q_{cap,t}^{DAC} = \beta_{DAC} \delta_{DAC,t} . P_{DAC}^{r} \quad (18)$$

$$u_{DAC(t)} \delta_{DAC,min} \leq \delta_{DAC(t)} \leq u_{DAC(t)} \delta_{DAC,max}, u_{DAC(t)} \in \{0,1\} \quad (19)$$

$$P_{E,t}^{DAC} = \varpi_{E} Q_{cap,t}^{DAC} \quad (20)$$

$$\Psi_{DAC,t}^{Absorp} = \Psi_{AB}^{DAC} Q_{cap,t}^{DAC} \quad (21)$$

$$QCO_{2,t}^{rel} = QCO_{2,t}^{total} - (Q_{co2(t)}^{pcc} + Q_{cap,t}^{DAC}) \quad (22)$$

In addition, the two forms of DAC (aqueous (DAC1) and solid sorbent (DAC2)) are analysed in this study. For DAC1, the thermal demand is provided by burning natural gas, and the consumption is computed in Eq. 24 where $\varpi_{NG}$ is the natural gas unit rate. Whereas the thermal demand when DAC2 is considered is provided by the waste heat generated within ZCMES, Eq. 25 estimated the thermal demand by DAC2 where $\varpi_{H}$ denotes the thermal demand unit of the system.

Aqueous solvent:

$$P_{NG,t}^{DAC} = \begin{cases} \varpi_{NG} QCO_{2,t}^{DAC,A}, u_{DAC(t)} = 1 \\ 0, u_{DAC(t)} = 0 \end{cases} \quad (24)$$

If a solid sorbent is used, which can utilise waste heat:

$$P_{H,t}^{DAC} = \begin{cases} \varpi_{H} QCO_{2,t}^{DAC,A}, u_{DAC(t)} = 1 \\ 0, u_{DAC(t)} = 0 \end{cases} \quad (25)$$

*2.5 Objective function of the proposed model*

The primary objective of optimal scheduling is to minimise the associated economic implication of the system dispatch. Hence, all possible economic costs are considered in this study. This comprises systems' operation costs ($C_{op,t}$), fuel cost ($C_{fuel,t}$), carbon capture and storage costs ($C_{CDR,t}$), and the environmental penalty cost ($C_{CO2,t}$). The adopted computational approach for each economic component is described in Eq. 26 - Eq. 29, and the total scheduling cost is computed in Eq. 30.

$$C_{op,t} = \alpha_{RES} P_{RES,t} + \alpha_{CFP} P_{CFP,t}^{e} + \alpha_{CFP} P_{GT,t}^{e} + \alpha_{WSHP} Q_{WSHP,t} + \alpha_{GB} Q_{GB,t} + \alpha_{EC} Q_{EC(t)}^{c} + \alpha_{AC} Q_{AC(t)}^{c} + \alpha_{BES} \left| P_{BES(t)} \right| + \alpha_{TES} \left| P_{TES(t)} \right| \quad (26)$$

$$C_{fuel,t} = \alpha_{gas} (\Pi_{gas,t}^{CHP} + \Pi_{gas,t}^{GB}) + \alpha_{coal} P_{coal,t} \quad (27)$$

$$C_{CDR,t} = \alpha_{CDR}^{CAP} \left( Q_{co2(t)}^{cap} + QCO_{2,t}^{DAC,A} \right) + \alpha_{CDR}^{STR} \left( Q_{co2(t)}^{cap} + QCO_{2,t}^{DAC,A} \right) + \alpha_{PCC}^{sorb} \Psi_{PCC,t}^{Absorp} + \alpha_{DAC}^{sorb} \Psi_{DAC,t}^{Absorp} + \alpha_{gas} P_{NG,t}^{DAC} \quad (28)$$

$$C_{CO2,t} = \alpha_{CO2} QCO_{2,t}^{rel} \quad (29)$$

$$\boldsymbol{F_t} = C_{op,t} + C_{fuel,t} + C_{CDR,t} + C_{CO2,t} \quad (30)$$

*2.6 CDRT proposed evaluation metrics*

Two evaluation metrics are proposed in this study to compare the considered CDRT configurations, these are $CO_2$ released indicator (CRI), and $CO_2$ captured released ratio (CCRR). CRI estimates the quantity of released $CO_2$ to the total emitted $CO_2$ by the MES. On the other hand, CCRR evaluates the ratio between captured $CO_2$ by the CDRT and the released $CO_2$. The mathematical expressions of the two metrics are shown below:

$$\text{CRI} = \frac{CO_{2,emit,t} - CO_{2,cap,t}}{CO_{2,emit,t}} \quad (31)$$

$$\mathrm{CCRR} = \frac{\mathrm{CO}_{2,\mathrm{cap},t} - \mathrm{CO}_{2,\mathrm{rel},t}}{\mathrm{CO}_{2,\mathrm{cap},t}} \quad (32)$$

$\mathrm{CO}_{2,\mathrm{emit},t}$: *emitted $CO_2$ by MES at each timestep*

$\mathrm{CO}_{2,\mathrm{cap},t}$: *captured $CO_2$ by MES at each timestep*

$\mathrm{CO}_{2,\mathrm{rel},t}$: *released $CO_2$ to the atmosphere at each timestep*

## 3.0 Proposed method

The proposed DRL method is grouped into three main categories: 1) Conversion of the problem into an RL-solvable task, 2) description of the adopted DRL algorithm structure, and 3) the training process of the algorithm that is integrated with a hyperparameter tuning feature. The methodological framework of this study is summarised in Fig. 5, and Eq. 6 illustrates the DRL interaction with the ZCMES environment.

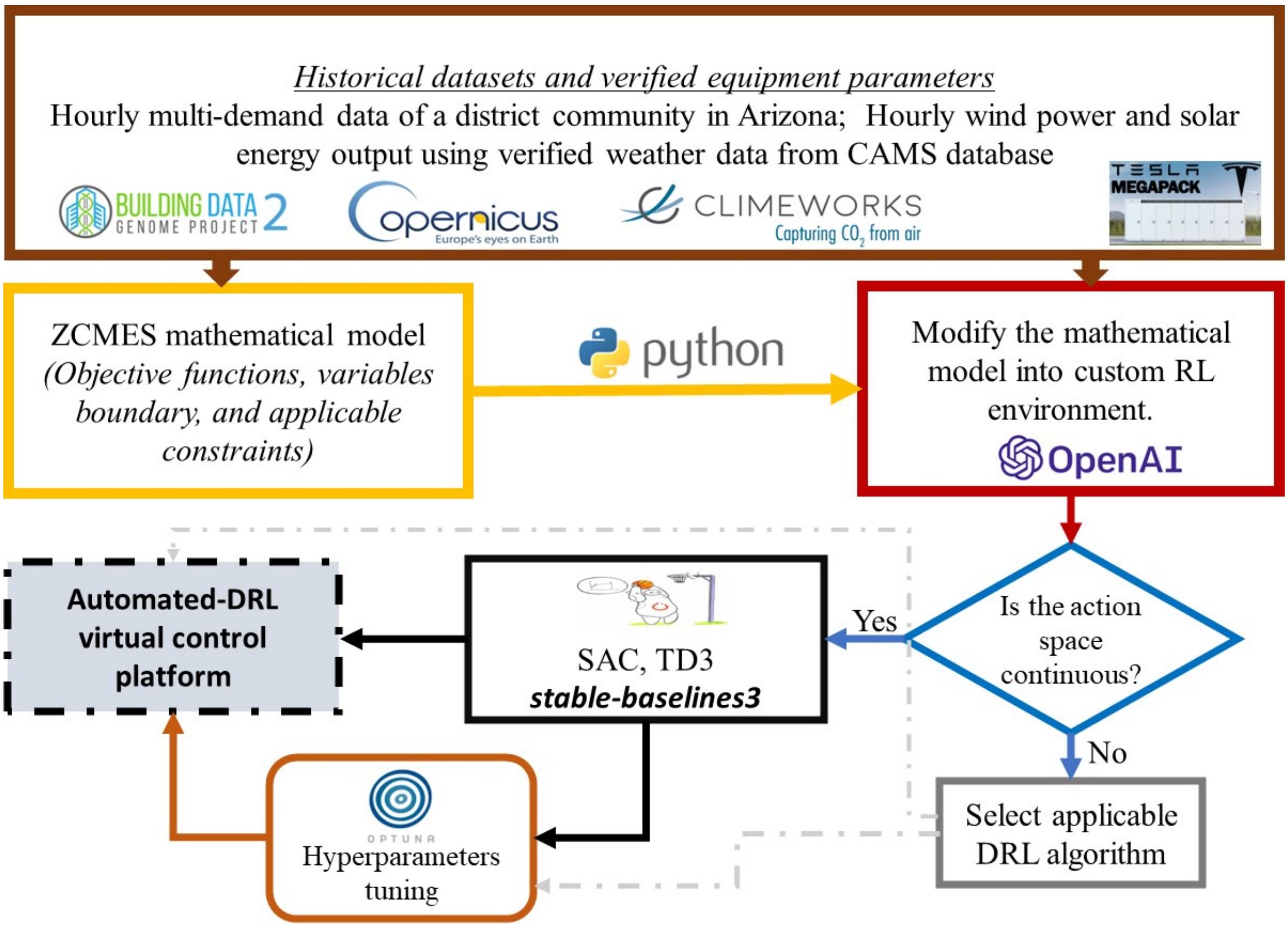

**Figure 5.** Proposed automated-DRL framework.

### *3.1 Scheduling problem conversion into a reinforcement learning task*

The first step in DRL applications is to specify the task as a Markov Decision Process (MDP). The core idea of an MDP is that the possible executed action(s) at the state $s_{t+1}$ only consider the state information at time step *t*. The RL framework in this study (Fig.6) comprises an agent and environment that interact iteratively during the offline training process. In this study, a virtual powerplant for autonomous decision-making is the agent, and the ZCMES is the environment. An MDP is expressed in a compact form as $MDP = (S, A, P, R, \gamma)$. The scheduling problem in this study is written as an MDP as follows:

1) *State space S*: In this study, the current state information, as shown in Fig 5, is the multi-energy demand of the prosumers, WT and PV generation output, and the current state of ZCMES that is needed in executing an action at the next time step.

   *State-space* = $[PV_t, PW_t, EL_t, HL_t, CL_t, SOCb_t, SOCH_t, P_{t,y}]$

2) *Action space A*: The DRL agent is responsible for sending a control signal $a_t$ to the ZCMES actuator to control the power output of each energy unit. In this study, the controlled actions by the DRL agent are the loading rate of each energy. The values are normalised between 0 and 1. This excludes energy storage systems between -1 and 1 to facilitate the charging and discharging. The agent selects an action based on its policy strategy $\pi$, which will be optimised during training by the selected DRL algorithm

$$a_t = [\delta_{RES(t),}\delta_{CHP(t)}, \delta_{GT(t)}, \delta_{GB(t)}, \delta_{HP(t)}, \delta_{BES(t)}, \delta_{TES(t)}, \delta_{EC(t)}, \delta_{AC(t)}, \delta_{PCC(t)}, \delta_{DAC(t)}]$$

3) *Reward function R*: The reward $r_t$ at each time step is the immediate reward the agent receives by executing the action $a_t$ based on the state information $s_t$. In this study, the reward value is taken as the cost function of the system, as shown in Eq. 26- Eq. 30. Notably, the system is subjected to some energy balance constraints and ramp limit constraints of the energy equipment to ensure the safety of the system. Hence, corresponding penalty functions are introduced as described below:

$$\Theta(EL_t) = \Theta_{el}\left(P_{RES\to EL,t} + P^{e}_{CFP\to EL,t} + +P^{e}_{GT\to EL,t} + P_{BES\to EL,t} - EL_t\right)^2 \tag{33}$$

$$\Theta(HL_t) = \Theta_{hl}\left(P^{h}_{CFP\to QL,t} + P^{h}_{GT\to QL,t} + P_{GB\to QL,t} + P_{WSHP\to QL,t} + P_{TES\to QL,t} - HL_t\right)^2 \tag{34}$$

$$\Theta(CL_t) = \Theta_{hl}\left(Q^{c}_{EC(t)} + Q^{c}_{AC(t)} - CL_t\right)^2 \tag{35}$$

$$\mathfrak{F}_{y,t} = \begin{cases} \omega^1|P_{t-1,y} - P_{t,y}|, P_{t-1,y} - P_{t,y} < -R \\ \omega^1|P_{t-1,y} - P_{t,y}|, P_{t-1,y} - P_{t,y} > R,\ y \in \{GT, CFP, GB, HP, EC, AC\} \\ 0, -R \le P_{t-1,y} - P_{t,y} \le R \end{cases} \tag{36}$$

$\Theta_{el}, \Theta_{hl}, \Theta_{hl}$ and $\omega^1$ are the penalty factors and $EL_t, HL_t,$ and $CL_t$ denote the electricity load, heat load, and cooling load. The ramp limit constraint is expressed in Eq. 36, where $R$ is the ramp limit value taken as 20% of the installed capacity and $P_{t,y}$ is the power output of the device $y$ at time $t$. The total penalty functions, including SOC constraints, are computed as follows:

$$\Theta_{PF} = \Theta(EL_t) + \Theta(HL_t) + \Theta(CL_t) + \psi_{x,t} + \mathfrak{F}_{y,t}, t = 1,2, \dots T, x \in \{BES, TES\} \quad (37)$$

Finally, the immediate reward at time slot $t$ is:

$$R_t = -(F_t + \Theta_{PF,t})\Delta t, \qquad t = 1,2, \dots T \quad (38)$$

***Transition probability P***: In this study, the transition probability denotes the influence of the agent's actions on the environment based on the current state information that serves as part of observation for the next state information. Strictly speaking, these are energy storages (BES and TES) $SOC_t$ and equipment energy output $P_{t,y}$.

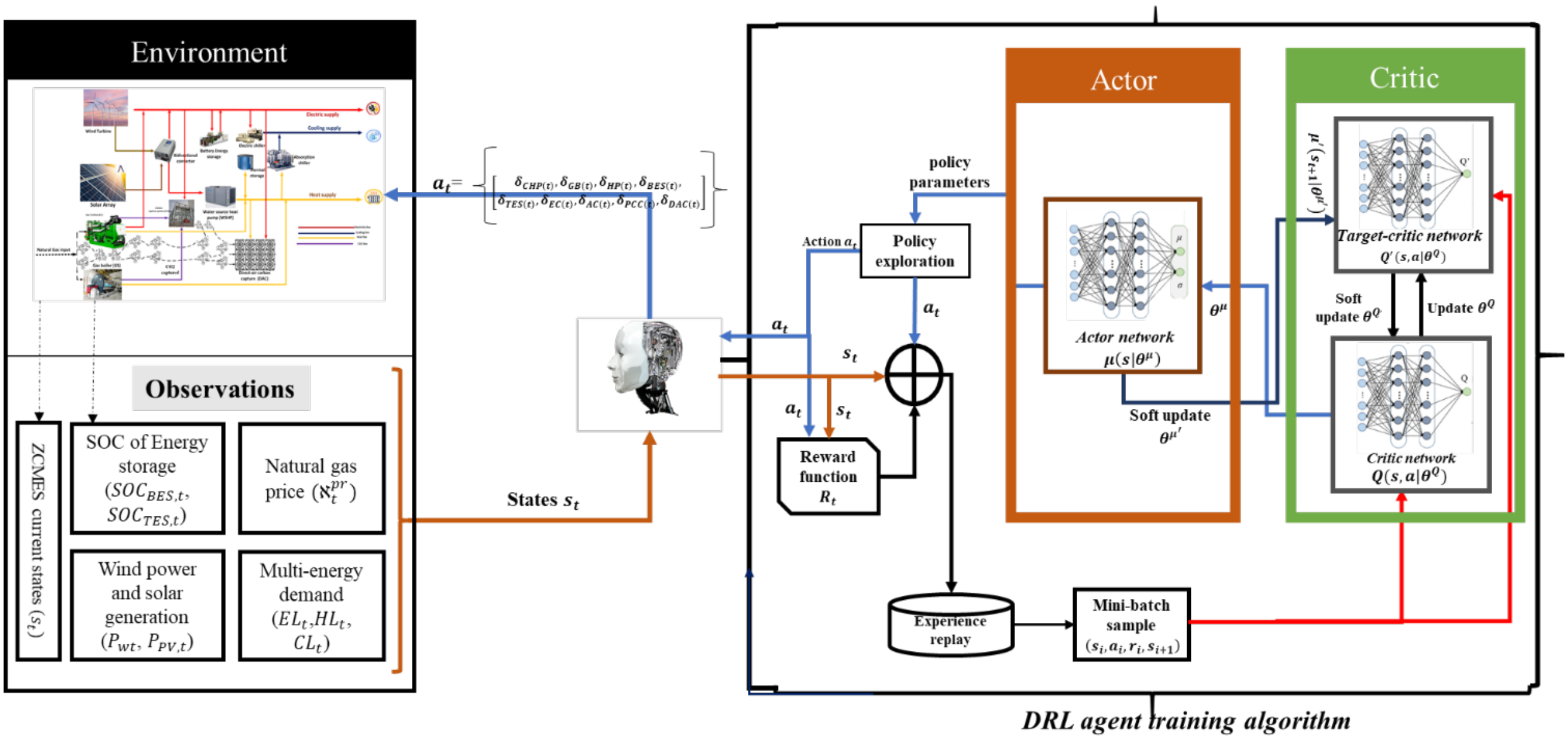

**Figure 6.** Interaction between DRL agent and ZCMES environment.

*3.2 Soft-actor (SAC) deep reinforcement learning algorithm*

The primary challenge of model-free DRL is high sampling complexity when handling high dimensional action, state spaces, and weak convergence. Deep deterministic policy gradient (DDPG) is the most adopted off-policy algorithm that addresses these challenges by improving sampling efficiency [35]. However, its performance is affected by hyperparameter tuning and its ineffective nature when the environment is affected by uncertainty [36]. The suitability of the DRL algorithm also depends on the nature of the action space, which can be continuous or discrete. Proximal policy optimisation (PPO) is a popular on-policy algorithm that applies to continuous state and action spaces. However, it suffers from sample inefficiency [37]. Hence, the soft actor-critic (SAC) algorithm that has improved sampling efficiency is proposed in this study [38]. Compared to DDPG, it possesses improved capability due to the introduction of maximum entropy to manipulate exploration and learning stability. Notably, it uses an actor-critic architecture and comprises actor and critic deep neural networks and a replay buffer for storing previous experience, as illustrated in Figure 7.

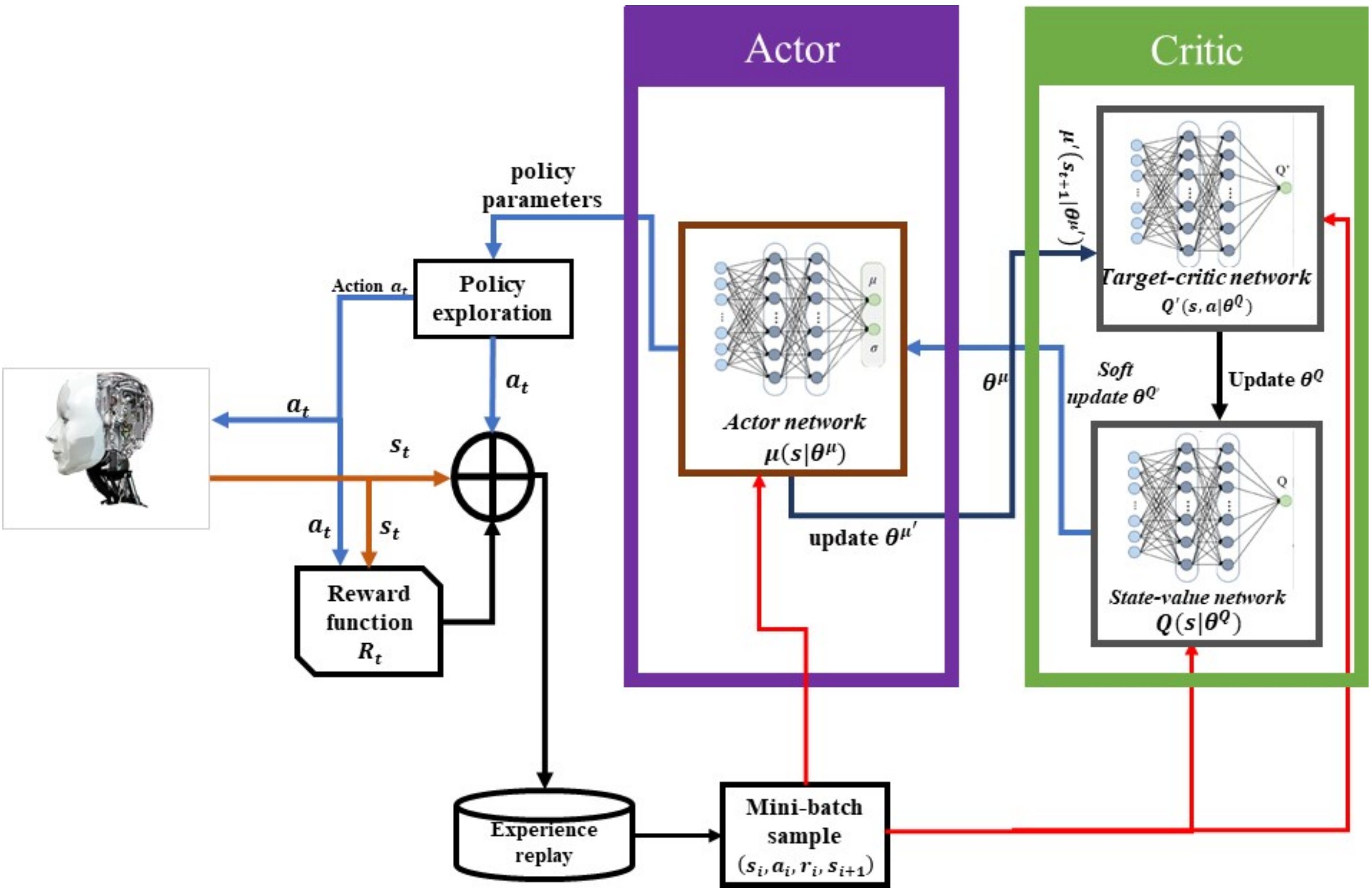

**Figure 7**. Soft-actor (SAC) deep reinforcement learning training process

*3.2.1 Soft-actor (SAC) algorithm network structure*

The core idea of SAC is to maximise the expected cumulative reward value while maximising the entropy simultaneously. Thus, the policy objective is shown in Eq. 39

$$J(\pi) = \sum_{t=0}^{T} E_{(s_t,a_t)\sim\rho_r}[r(s_t,a_t) + \alpha H(.|s_t))] \tag{39}$$

$$H(X) = -\sum_{x_i \in X} P(x_i) \log P(x_i) \tag{40}$$

Where $s_t$ and $a_t$ are the state and action information at time slot $t$, $r(s_t, a_t)$ denotes the immediate reward function of action $a_t$ at state $s_t$, $H(.)$ is the entropy incorporated to encourage policy exploration as computed in Eq. 40, and $\alpha$ is the temperature parameter introduced to regulate the policy's stochastic degree and control the entropy term's relevance against the reward. Since the aim of the DRL algorithm is to select the optimal policy that maximises the cumulative long-term reward, the optimal policy strategy is evaluated as:

$$\pi^*_{max} = argmax_{\pi} \sum_{t=0}^{T} E_{(s_t,a_t)\sim\rho_r}[r(s_t, a_t) + \alpha H(.|s_x))] \tag{41}$$

*3.2.2.1 Critic network*

The SAC algorithm consists of an actor and a critic network. The critic network is responsible for policy evaluation using a deep neural network (DNN) parameterised by $\theta$ to approximate the value function. Firstly, the output value of the critic network is used to estimate the soft Q-value $Q_\theta(s, a)$ utilising the randomly batched sample from the replay buffer, which houses the historical experience of agent interaction with the environment and its action at each timestep. Secondly, the performance evaluation of the DNN is estimated by obtaining the mean square error (MSE) between the estimated Q value and the target Q-value $\hat{Q}(s_t, a_t)$ as described below:

$$J_Q(\theta) = E_{(s,\alpha)\sim D}\left[\frac{1}{2}\left(Q_\theta(s_t, a_t) - \hat{Q}(s_t, a_t)\right)^2\right] \tag{42}$$

$$\hat{Q}(s_t, a_t) = r(s_t, a_t) + \gamma E s_{s,1}\sim p\left[V_{\overline{\varphi}}(s_{t+1})\right] \tag{43}$$

Where $V_{\overline{\varphi}}(s_{t+1})$ is the target state value, and $\gamma$ is the discount factor. To ensure the accuracy of the $Q_\theta(s, a)$, its parameter $\theta$ is updated by applying the gradient descent method on $J_Q(\theta)$:

$$\theta_{t+1} \leftarrow \theta_t - \alpha_c \nabla_\theta J_Q(\theta) \tag{44}$$

Where $\alpha_c$ denotes the learning rate of the critic network and $\nabla_\theta J_Q(\theta)$ is the gradient of $J_Q(\theta)$. Further, the training stability of the critic network is achieved by introducing a state-value function parameterised by $\varphi$. The network is updated by estimating its MSE, followed by its gradient estimate and parameter update in Eq. 46 and 47.

$$J_v(\varphi) = E s_{r\sim D}\left[\frac{1}{2}\left(V_\varphi(s_t) - E_{\alpha_\alpha\sim\pi_\emptyset}[Q_\theta(s_t, a_t) - \log \pi_\emptyset(\alpha_t|s_t)]\right)^2\right] \tag{45}$$

$$\varphi_{t+1} \leftarrow \varphi_t - \alpha_c \nabla_\varphi J_Q(\varphi) \tag{46}$$

Additionally, the parameter updating process of the critic is stabilised by applying a DNN with parameter $\hat{\varphi}$ to estimate the target state-value function $\widehat{V_\varphi}(s)$ follow by the parameter's update by applying a soft update on the state-value network weights:

$$\widehat{\varphi_{t+1}} \leftarrow m\varphi_{t+1} + (1 - m)\widehat{\varphi_t} \tag{47}$$

Where $m \in (0,1)$ is the smoothing factor.

*3.2.2.2 Actor network*

The primary aim of an actor network is policy improvement. Hence, a DNN approximator function that is parameterised with $\psi$ to formulate policy $\pi_\psi(.|s)$ is trained following Eq. 42. Due to the intractability of sampling EBP model like Gaussian distribution, a state-conditioned stochastic network is applied to sample the action followed by KL divergence to evaluate the difference between $\pi_\psi^*(.|s)$ and $\pi_\psi(.|s)$.

$$L(\psi) = E\left[D_{KL}\pi_\psi(.|s)||\pi_\psi^*(.|s)\right] \quad (48)$$

To reinforce the actions against states' uncertainties influence, a reparameterisation trick $f_\emptyset(\varepsilon_t; s_t)$ is introduced where $\varepsilon_t$ is the standard normal distribution sample that generates the noise signal [39]. Explicitly, for each state $(s_t)$ intake of policy $\pi_\psi(.|s)$, it output mean $\pi_\psi^\mu(s_t)$ and standard deviation $\pi_\psi^\sigma(s_t)$ that correspond to the decision-making action of the actor, as shown in Eq. 50.

$$a_t = f_\emptyset(\varepsilon_t; s_t) = \pi_\psi^\mu(s_t) + \varepsilon_t \pi_\psi^\sigma(s_t) \quad (49)$$

Eq. 48 is re-written explicitly with the introduction of Eq. 49 as shown in Eq. 50. Further, to update the actor parameter $\psi$, the gradient of $L(\psi)$ is evaluated, followed by the parameter update as described in Eq. 51 where $\alpha_a$ denotes the actor's learning rate, which is a small positive value.

$$L(\psi) = E_{s_t \sim D, \varepsilon_t \sim N(0,1)}[\log \pi_\emptyset(f_\emptyset(\varepsilon_t; s_t)|s_t) - Q_\theta(s_t, f_\emptyset(\varepsilon_t; s_t))] \quad (50)$$

$$\psi_{t+1} \leftarrow \psi_t - \alpha_c \nabla_\psi L(\psi) \quad (51)$$

*3.3 DRL training and parameters updating process*

As stated in the preceding section, SAC comprises three main components: Actor-network, critic network, and experience replay buffer. The networks are trained using stochastic gradient descent. During the training process, the past experiences $e_i = (s_t, a_t, r_t, s_{t+1})$ are stored in the replay buffer to form a tuple $M = \{e_1, e_2 \dots e_M\}$. The agent uses a mini-batch of experience randomly sampled from the buffer to update the actor and critic network at each learning process. Furthermore, for the selected actions by TD3, a stochastic noise model is added to cater for the stochasticity of the observation states. In contrast, the SAC algorithm uses a stochastic policy for action mapping. After several iterations, the learning process turns out after reaching the final state or the end of the episode.

| SAC algorithm Training process |
|---|
| 1: Initialise actor network $\mu_{\emptyset}$, critic network $Q_{\theta}$ with random weights $\emptyset\ and\ \theta$ |
| 2: Initialise the associated target networks with weights $\emptyset'\ and\ \theta'$ |
| 3: Initialise a replay buffer D and timestep T |
| 4: for Timestep = 1: T **do** |
| 5: receive the initial state $s_t$ from the ZCMES environment |
| 6: DRL agent selects action $a_t = \mu_{\emptyset}(s_t)$ using stochastic policy according to the current state $s_t$ |
| 7: apply the action $a_t$ on the environment, observed the reward $r_t$, and the new state $s_{t+1}$ |
| 8: store experience $(s_t, a_t, r_t, s_{t+1})$ in replay buffer D |
| 9: sample a random minibatch of N experiences from the replay buffer D |
| 10: update the actor, critic, and associated target networks as described in the supplementary materials. |
| 11: update the environment state $s_t \leftarrow s_{t+1}$ |
| 12: **end for** |

**4.0 Case study**

The success of the DRL algorithm depends on the availability and accessibility of rich historical data and the proposed systems' validated parameters. Thus, a year multi-energy load and the weather data between January 1, 2017, and December 31, 2017, of a district community in Arizona, United States, is used in this study (Fig. 8). The data is in hourly resolution and was retrieved from the Building Data Genome Project 2 (BDGPII) database [40]. Further, the solar radiation data wind speed measurement for the estimation of renewable power during this period was obtained from the Copernicus Atmosphere Monitoring Services (CAMS) website [41].

Since coal continues to play an important role in sustaining the electricity grid, the feasibility of a coal-fired plant in a modular compartment has become one of the objectives of the U.S. Department of Energy under the Advanced Coal Energy (ACE) program [42]. Hence, we considered a medium-scale modular coal-fired plant economical for a district community in our model. The installable renewable energy system is restricted to the available rooftop area (127km$^2$) in the community for PV installations and a maximum of 100nos of WT, considering space constraints. For DAC, the Climeworks DAC technology specification is adopted in this study for the sorbent-based DAC. This system has a capturing efficiency of 90% and can utilise waste heat for regeneration.

Moreover, the sorbent consumption is 3kg for capturing 1 ton of $CO_2$ [23]. On the other hand, the specifications provided by Keith et al. [43] after a large-scale pilot study on industrial DAC were used for the solvent-based DAC. The capturing efficiency of the system is 85% and requires 5.25GJ of natural gas and 366kWh of electricity for capturing 1 ton of $CO_2$. For the PCC, solvent-based $CO_2$ removal is considered due to its maturity following the specification of the Petra Nova pilot study project with 90% removal efficiency and high-performance solvent [44]. This study adopts the Amine-based solution MEA for both PCC and solvent-based DAC. The absorption capacity of the solution is taken as 53.68kg/ton of captured $CO_2$, which is obtained from [45] based on a comprehensive simulation that is validated by pilot plant experiment results, which comprises MEA evaporation in the absorber (53.4) and the loss after water wash in the slaker (0.28kg/ton). In addition, the carbon capture and storage cost are taken as 50$/ton and 10$/ton, respectively [46], while the solid sorbent and solvent costs are taken as10$/kg and 3$/kg [47], respectively.

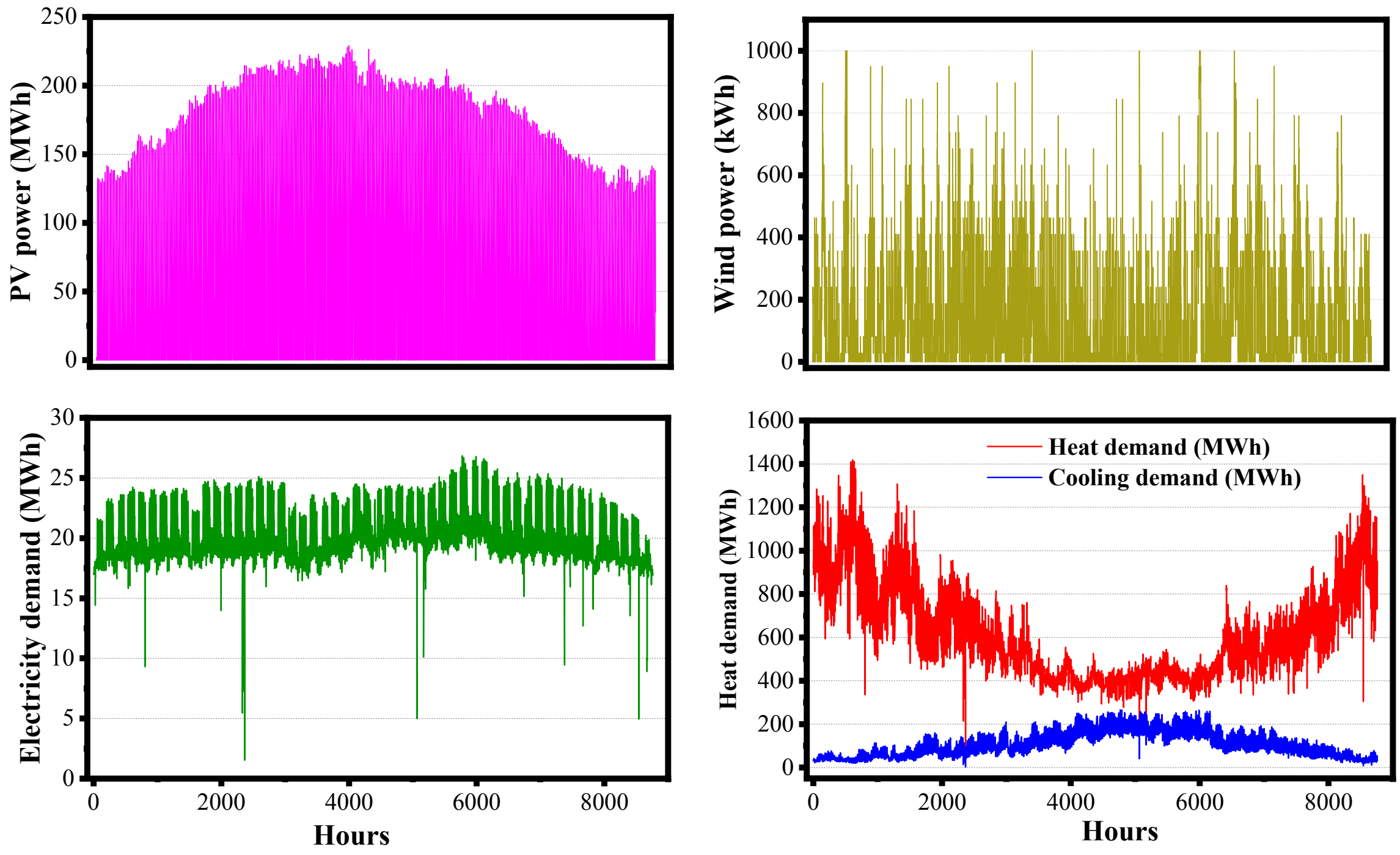

**Figure 8**. Hourly multi-energy data and renewable power of a district community in the studied year.

### 4.1 Simulation results and discussions

*4.2 Automated hyperparameter tuning of DRL algorithm*

This study introduces an automated hyperparameter tuning feature for any chosen DRL algorithm. Since deep neural networks are function approximators, the selection of each hyperparameter value determines the algorithm's optimality and convergence. Conventionally, selecting appropriate hyperparameter values is a daunting task for researchers. Hence, a twin-delayed DDPG (TD3), an improved version of DDPG [48] and SAC, are integrated with hyperparameter tuning features and evaluated in the current study.

Thus, the Optuna module [49] is integrated with the DRL algorithm for automatic hyperparameter tuning. The main procedure is to specify the various range of each hyperparameter, and then Optuna selects suitable combinations that maximise the defined objectives after a specified number of trials. In this study, the evaluated policy mean reward for each DRL algorithm is specified as the objective function to be maximised. The trial period is set to 50 with the integration of the Optuna pruner feature to trim undesiring avenues. A custom call-back function is introduced to terminate the training process if the difference between objective function values is lower than a 0.0005 threshold after 50 trials. The timestep is set to 30,000 at 1hour resolution for each timestep. Fig. 9 illustrates the importance of each hyperparameter and the optimal objective value of each DRL agent obtained while the selected values are presented in Table 1. Batch size is ranked first as the most important feature of the SAC algorithm, followed by the smoothing factor and the network's learning rate. On the other hand, the activation function is the main influencing factor of TD3 (twin delayed DDPG) performance, followed by the cumulative reward discount factor (gamma) and batch size.

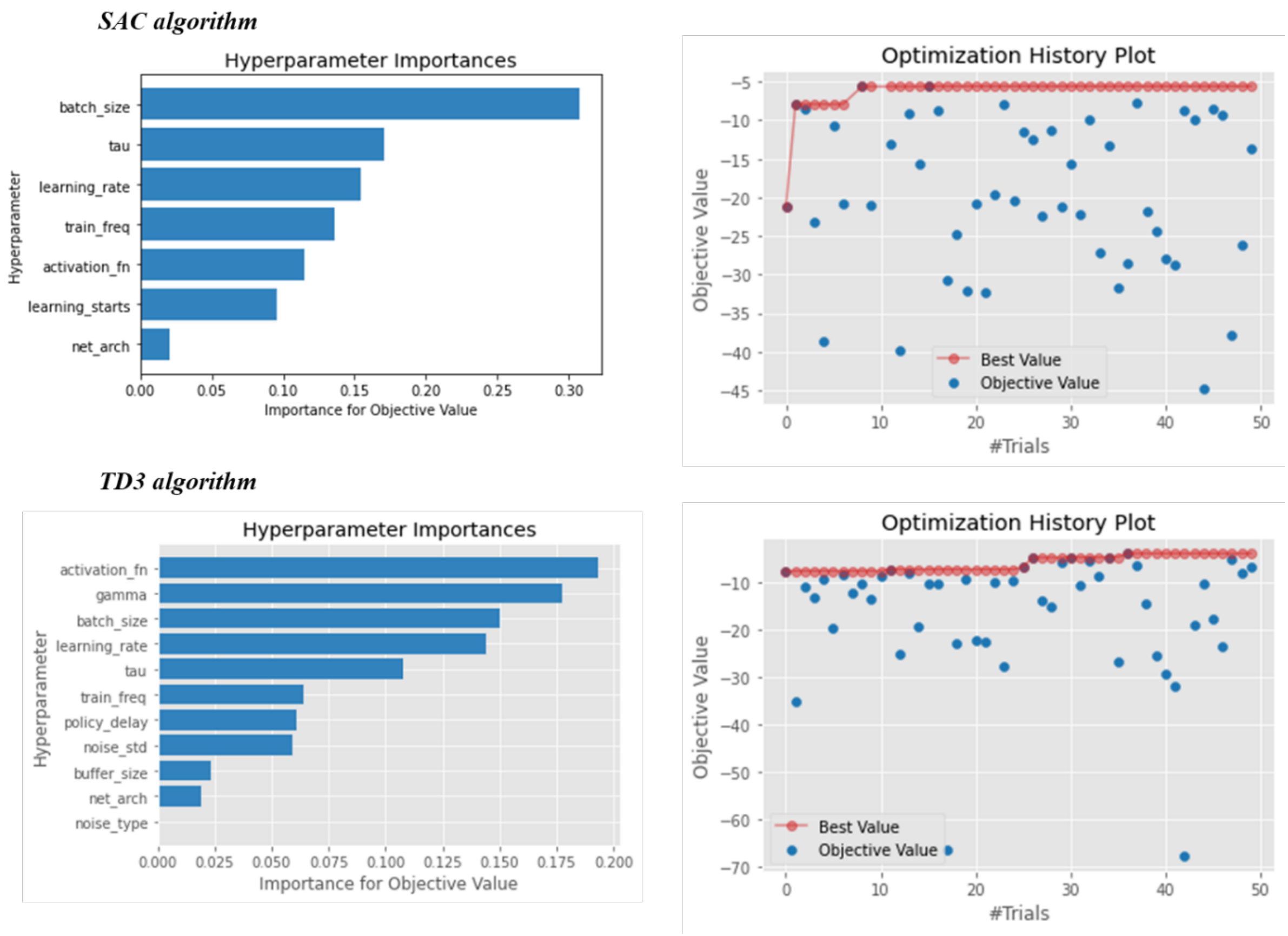

**Figure 9.** Hyperparameters ranks and optimisation plot obtained by Optuna.

**Table 1.** Selected hyperparameters by Optuna.

| | **SAC** | **TD3** |
|---|---|---|
| gamma | 0.96 | 0.90 |
| Learning rate | 0.00122 | 0.00148 |
| Buffer size | 100,000 | 100,000 |
| Batch size | 512 | 1024 |
| Learning starts | 1000 | - |
| Training frequency | 1 | 1 |
| Tau (smooth factor) | 0.02 | 0.08 |
| Noise type | - | normal |
| Entropy coefficient. | 0.05 | - |
| Noise-std | - | 0.5237 |
| Net-arch | {400, 300} | {400, 300} |
| **Mean-reward** | **-5.59** | **-3.91** |

The performance of the considered DRL algorithms during offline training is shown in Figure 10. The visualisation shows that the algorithms with default hyperparameters fail to converge after the total timesteps. With the hyperparameter tuning feature, a better convergence speed from 30,000 timesteps is achieved by both SAC and TD3. However, there is an interesting observation as SAC yielded optimal mean reward compared to TD3; the reason for this is its maximum entropy feature that enables improved exploration.

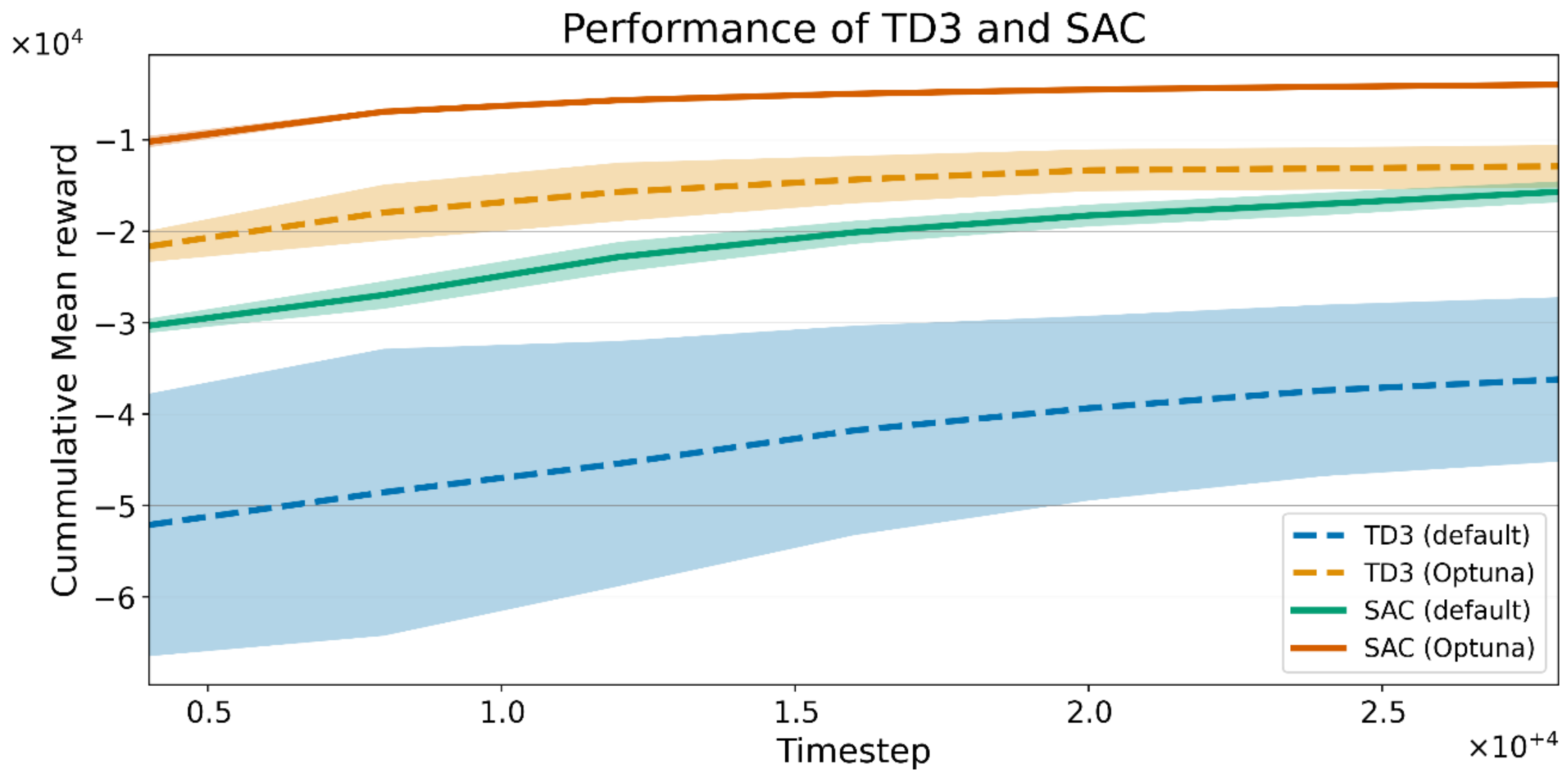

**Figure 10.** Performance comparison of SAC and TD3.

*4.3 CDR technology influence*

The two prominent CDRTs are evaluated in this study. We consider four (4) possible cases to achieve this; Case 1 serves as the baseline without CDRT. The possibility of PCCS as CDRT is considered in Case 2. We propose a combination of PCCS and a solvent-based DAC in Case 3, while sorbent-based DAC, also known as low-temperature DAC, is considered in Case 4. Since our focus in this study is to develop a DRL agent for autonomous scheduling, we created a SAC agent for each case study using the obtained hyperparameter values by Optuna in the preceding section. The proposed agents are trained for a total time step of 100,000, and Figure 11 illustrates the agents' performance. All the agents converged at the end of the training. There is a difference among the agents' cumulative reward due to variations in the number of actions executed by each agent and the stochastic policy nature of the SAC algorithm. After successful offline training, the agents are ready for deployment in real-time in an online mode, and a typical day is selected for further analysis in the following section.

*Scenario description*

**Case 1 (Agent 1):** Optimal dispatch strategy without CDR technology (baseline model)

**Case 2 (Agent 2):** Optimal dispatch strategy with PCCS

**Case 3 (Agent 3):** Optimal dispatch strategy with PCCS and DAC1 (solvent-based DAC)

**Case 4 (Agent 4):** Optimal dispatch strategy with PCCS and DAC2 (sorbent-based DAC)

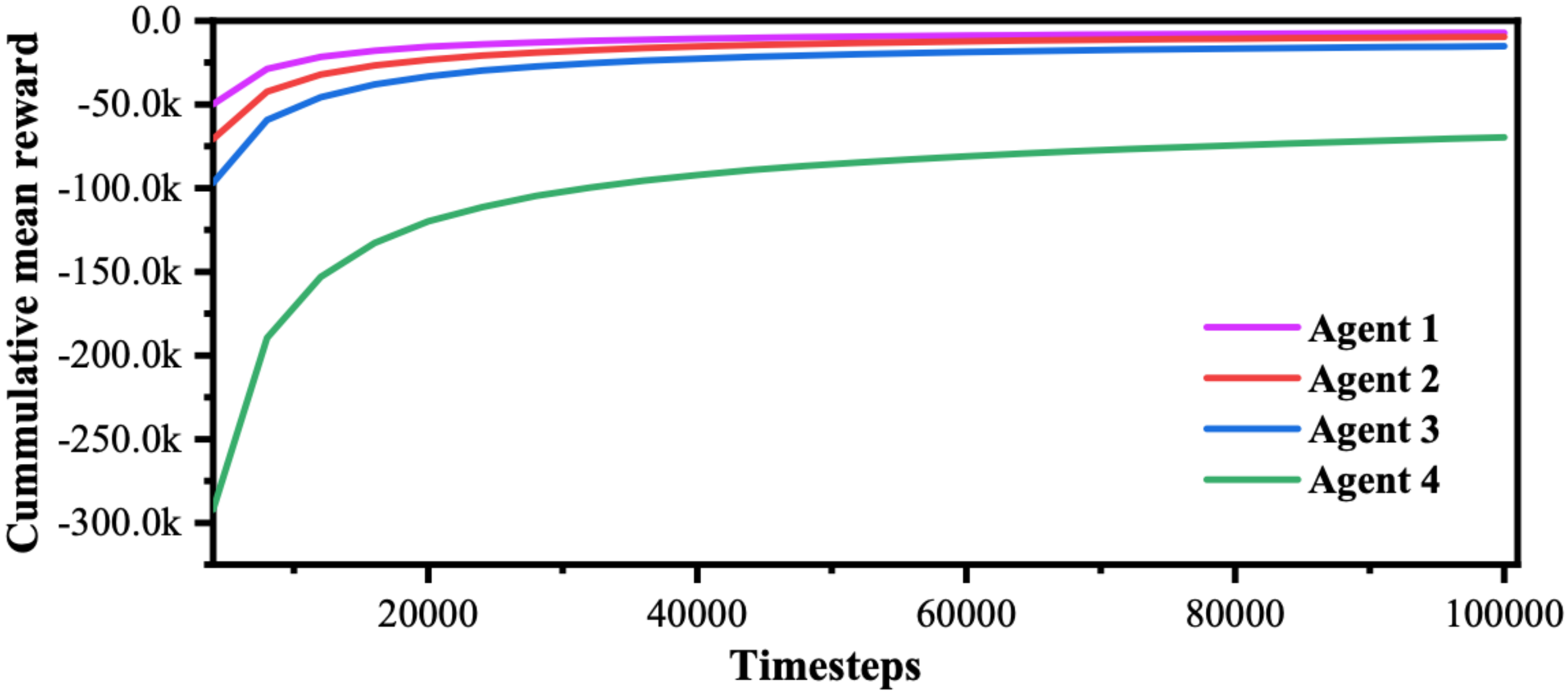

**Figure 11.** Proposed SAC DRL agents' performance during training.

*4.4 Multi-energy demand provision scheduling by the trained DRL agent*

Meeting the multi-energy demand of the prosumers without frequency imbalance or power mismatch is the primary aim of any interconnected MES. Hence, our first step in this study is to test the scheduling actions of the trained DRL agent during online deployment. Figure 12 illustrates the multi-energy distribution schedule by Agent 4, which is the most complex environment. The electricity demand (EL) of the prosumers is sufficiently met at each timestep as illustrated in Fig. 12a. At the same time, the remaining electrical power is supplied to other energy converters within MES or stored in the BES optimally. Specifically, the electrical demand of EC (Pec) and WSHP (Php) to generate cool air and heat when needed, the power demand of CDR (i.e., PCCel and DACel), and the charging of BES. It is worth stating that priority is given to the total utilisation of the available renewable energy by PV and WT to avoid extra costs on curtailment before considering the CHP power generation (Gte and Pcoale). Figure 12b and 12c illustrates the thermal demand-supply balance by the proposed DRL agent. The thermal (heat and cold) demand is also adequately met, including the provision of heat demand for DAC regeneration purpose. Also, it can be seen that the cooling load is mainly provided by an absorption chiller (AC) throughout the day, while EC only complements the required cool air between 9:00 am and 12:00 noon. The reason for this is the vast availability of waste heat by the cogeneration plant that can be converted to cooling energy by AC. Lastly, Fig. 12d describes the SOC of each energy storage system (BES and TES). This gives information on the current state of the energy storage system, the evaluation of its usage, and the current energy level of the system. It can be observed that both storage systems SOC were at the minimum limit between 7:00 pm and 9:00 pm, which shows that there is a peak multi-energy demand during this period.

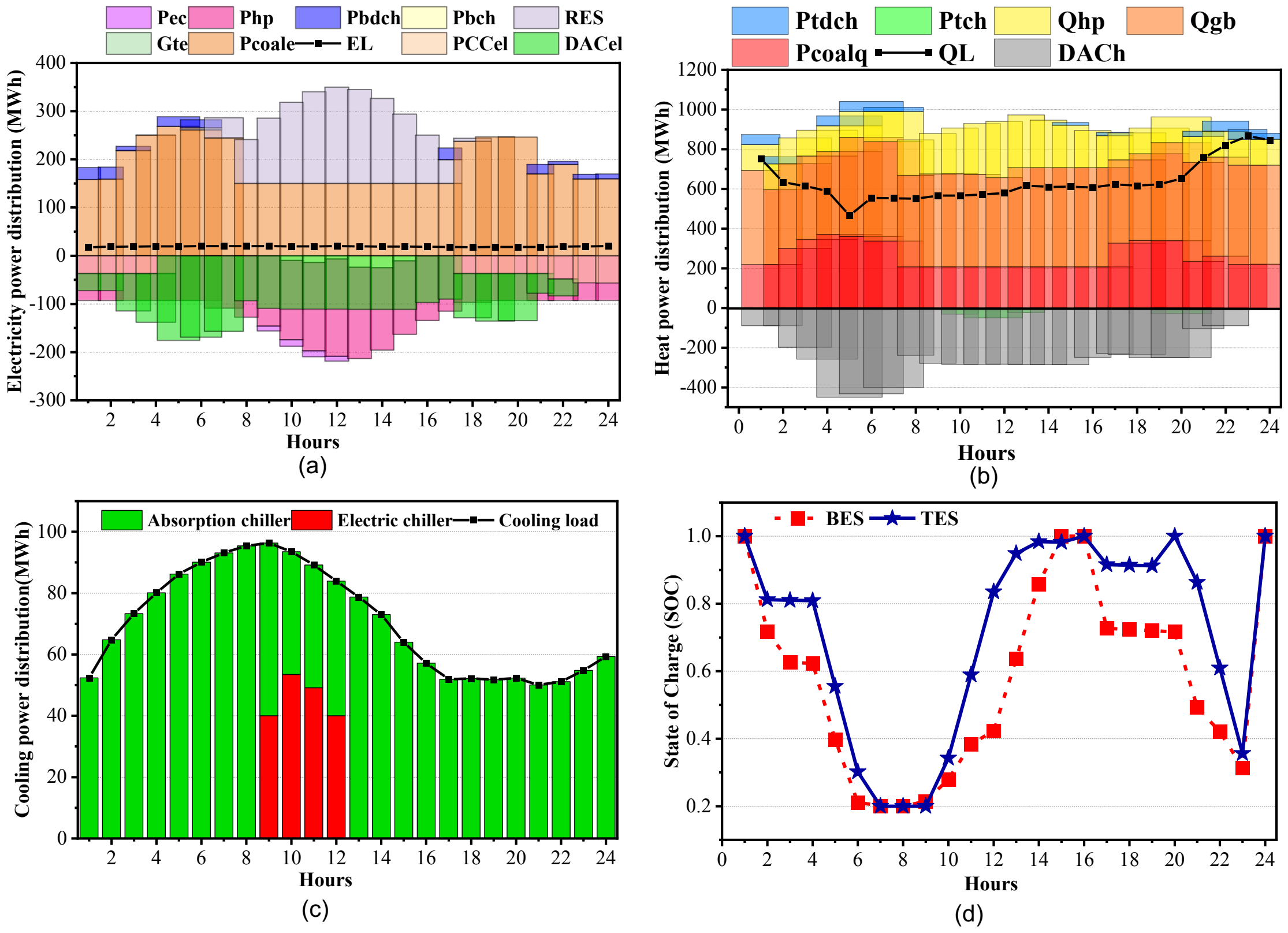

**Figure 12**. Multi-energy distribution by the trained DRL agent (Agent 4).

*4.5 Emission captured and released evaluation*

Figure 13 describes the $CO_2$ capturing performance of the trained DRL agents. A zero-carbon release was achieved between 9:00 and 15:00 when the integrated CDRT captured all the generated $CO_2$ during these periods. Specifically, Agent 3 and Agent 4 scheduled the DAC to capture a large portion of the generated $CO_2$. The figure also shows that emitted $CO_2$ by the MES varies with the CDRT configuration, which is influenced by the energy consumption pattern of the chosen CDRT. To aid the optimal selection of CDRT configuration with MES, we proposed two evaluation metrics which are $CO_2$ released indicator (CRI) and $CO_2$ captured-released ratio (CCRR).

The CRI estimates the quantity of released $CO_2$ to the total emitted $CO_2$ by the IES. On the other hand, CCRR evaluates the ratio between captured $CO_2$ by the CDRT and the released $CO_2$. Notably, an optimal CDRT configuration is expected to have a low CRI value and a high CCRR compared with other configurations. Figure 13d compares the CRI and CCRR of the trained agents. Agent 4 has the least CRI value of 2.53 and the highest CCRR value of 38.54 compared to other Agents. The reason for this is based on the optimal energy consumption by Agent 4 CDRT, especially its low-temperature heat demand that can be met by waste heat and the absorption capacity of the selected sorbent.

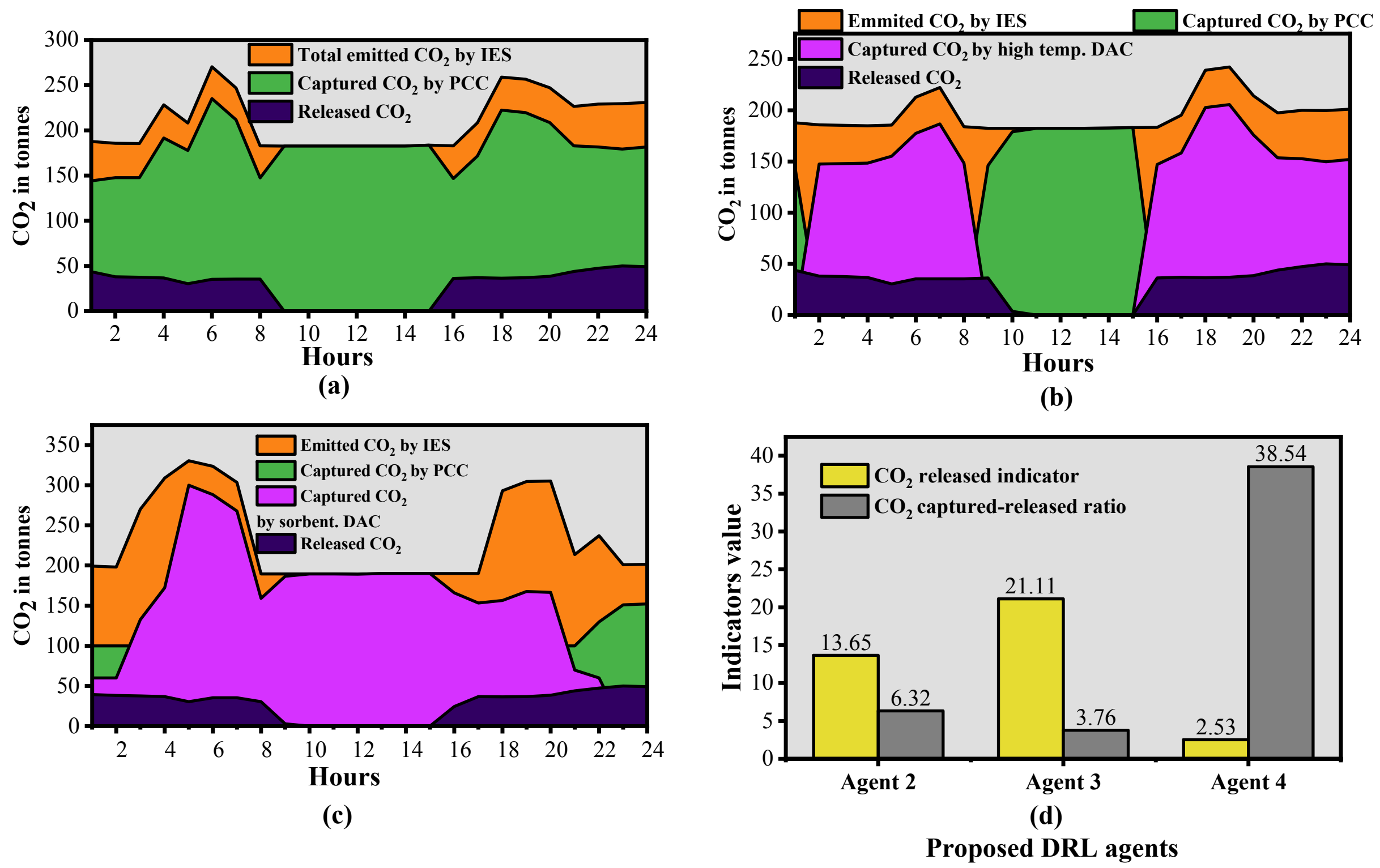

**Figure 13.** CDR performance of the trained DRL agents.

*4.6 CDR energy and materials schedule by the trained DRL agents*

The energy and solvent/sorbent consumption of the considered CDRT configuration is illustrated in Fig 14. The electricity consumption of PCC for all DRL agents is compared in Fig 14a. It is worth mentioning that the considered PCC in this study is integrated with an electric heat pump to meet the heat demand of the system. Thus, part of the electricity consumption is used to meet the thermal demand. Agent 2 has the highest PCC electrical consumption since only PCC was responsible for $CO_2$ capture.

In contrast, the burden on PCC reduces with the integration of DAC in Agent 3 and Agent 4. Meanwhile, DAC requires both electricity and thermal demand, as illustrated in Fig. 14b. The thermal demand by Agent 3 DAC is of high temperature that is generated by burning natural gas, while Agent 4 utilises the waste heat and the heat generated within IES to meet its thermal demand. Consequentially, the solvent and sorbent consumption for both PCC and DAC is described in Fig 14c and 14d. Agent 4 has the least PCC solvent and DAC sorbent consumption due to the DAC's high absorption sorbent capacity. It is worth stating that high solvent/sorbent consumption does not transpire in high $CO_2$ capturing rate, as it is dependent on the chemical properties of the selected solvent/sorbent. In addition, the unit cost of a solid sorbent is 70% higher than an aqueous solvent. However, the trained DRL agent still considers it to be economical considering the overall economic implication in the long run with the use of available thermal energy within MES.

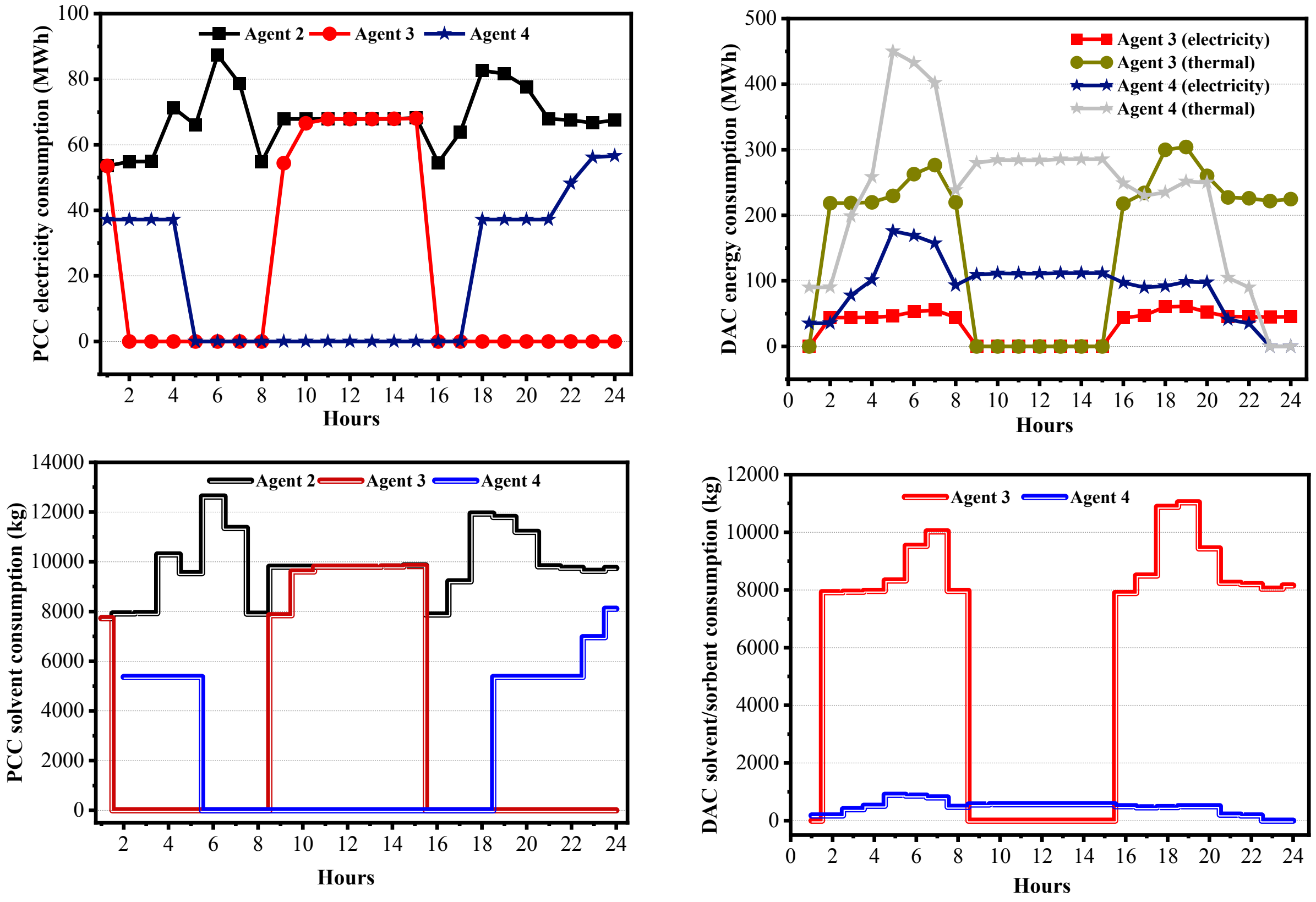

**Figure 14.** CDRT energy and solvent consumption.

*4.7 CDR technology's economic influence*

The economic backlash has always been the primary criticism of CDR technology. While this point may be valid from an economic perspective, Parmenides, an ancient Greek philosopher, once said, "nothing goes for nothing"; the environmental value must not be viewed only through an economic lens. However, the economic sacrifice must be paid optimally. In this regard, we evaluate the economic influence of CDRT when integrated with MES, as shown in Table 2. The baseline model, Agent 1, achieves the least overall cost as expected. The consideration of PCC by Agent 2 reduced emission costs by 75.35% and severely reduced the operation and fuel costs by utilising the full potential of the energy storage.

Nevertheless, investors will be cynical about the huge cost requirement for operating PCC. A solvent-based DAC is considered in Agent 3 to complement PCC. However, a 7.28% reduction in the CDR cost is achieved using the approach. Another challenge is the additional external input in the form of natural gas to meet the regenerator's high-temperature demand. Agent 4 handled the proposed configuration model in this study; with solvent-based DAC to complement PCC, the model achieved the least CDR cost with 36.53% and 31.54% reduction compared to Agent 2 and Agent 3, respectively. This cost reduction is attributed to the use of low-temperature waste heat that is generated by MES and the high absorption capacity of solid sorbent compared to an aqueous solvent.

**Table 2**. Proposed DRL agents' economic cost.

| USD | **Agent 1** | **Agent 2** | **Agent 3** | **Agent 4** |
|---|---|---|---|---|
| Operation cost | 191,503.59 | 55,554.40 | 55,058.59 | 72,597.88 |
| Fuel cost | 198,773.19 | 89,764.74 | 86,142.95 | 117,041.93 |
| Emission cost | 86,924.98 | 21,423.99 | 22,618.75 | 21,460.71 |
| CDR cost | - | 998,895.84 | 926,094.87 | 634,002.64 |
| Overall cost | **327,093.12** | **1,165,638.98** | **1,089,915.16** | **844,080.48** |

The economic components of each DRL agent's CDRT are illustrated in Fig. 15. To provide the energy demand (both electrical and thermal) by CDRT, a large chunk of the operation cost is allocated to solvent/sorbent cost. Specifically, 45.4% of the cost is consumed by DAC solvent, which depicts that DAC is scheduled to capture most of the generated $CO_2$. On the other hand, DAC sorbent contributes the least to the CDR cost by Agent 4. The reason for this is ascribed to the high absorption capacity of the sorbent despite its high unit cost of 10 USD/kg compared to an aqueous solvent of 3 USD/kg.

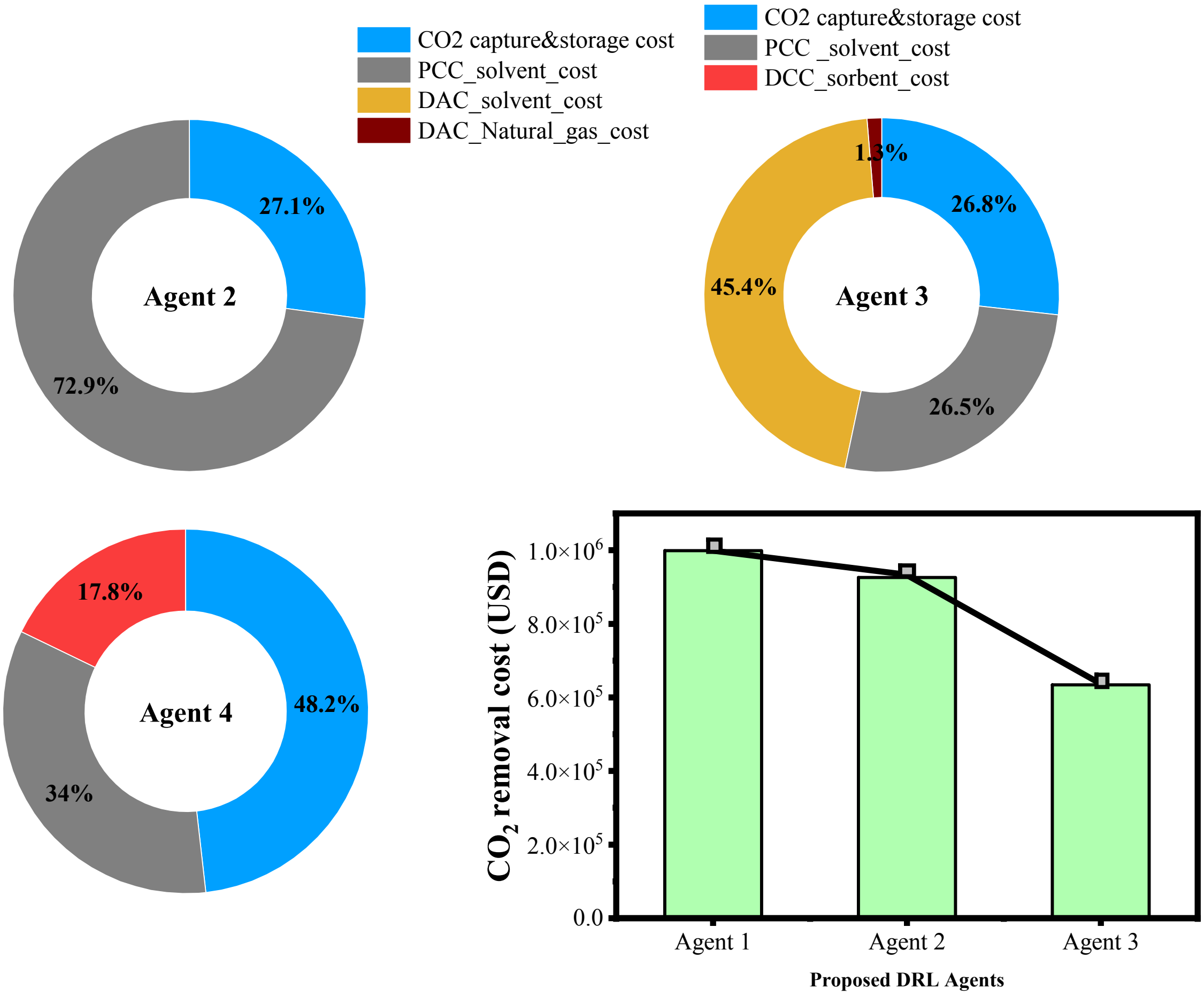

**Figure 15.** CDRT economic components.

*4.8 Carbon price influence on CDRT performance*

As stated in section 2.3.2, the primary criticism of CDRT is the enormous energy demand, if not properly scheduled, and the economic implication of the system when compared to other available options. Another perspective is that investors will pay for the carbon penalty cost, which is currently at 30USD-50USD/ton, instead of considering CDRT. Meanwhile, a monetary penalty should not be the primary yardstick to achieve the zero-carbon target. Therefore, we evaluate the influence of carbon price on CDRT performance in terms of the captured and released $CO_2$, as shown in Fig 16. The figure illustrates that the current carbon price does not encourage the adoption of CDRT. The captured/released $CO_2$ is achieved due to the restriction in the DRL agent environment, which stipulates that the maximum released $CO_2$ to the environment must not be more than 20% external grid emission. However, from 400USD/ton, the capturing rate of CDRT increases while the released $CO_2$ reduces. This shows that at the current economic cost of CDRT and the current energy, thermal, and chemical properties of its available technology, CDRT adoption is worthwhile from an economic point of view if the carbon penalty price is pegged at a minimum of 400USD/ton. In fact, zero-carbon is feasible at 450-500USD/ton, as shown in Fig 16b.

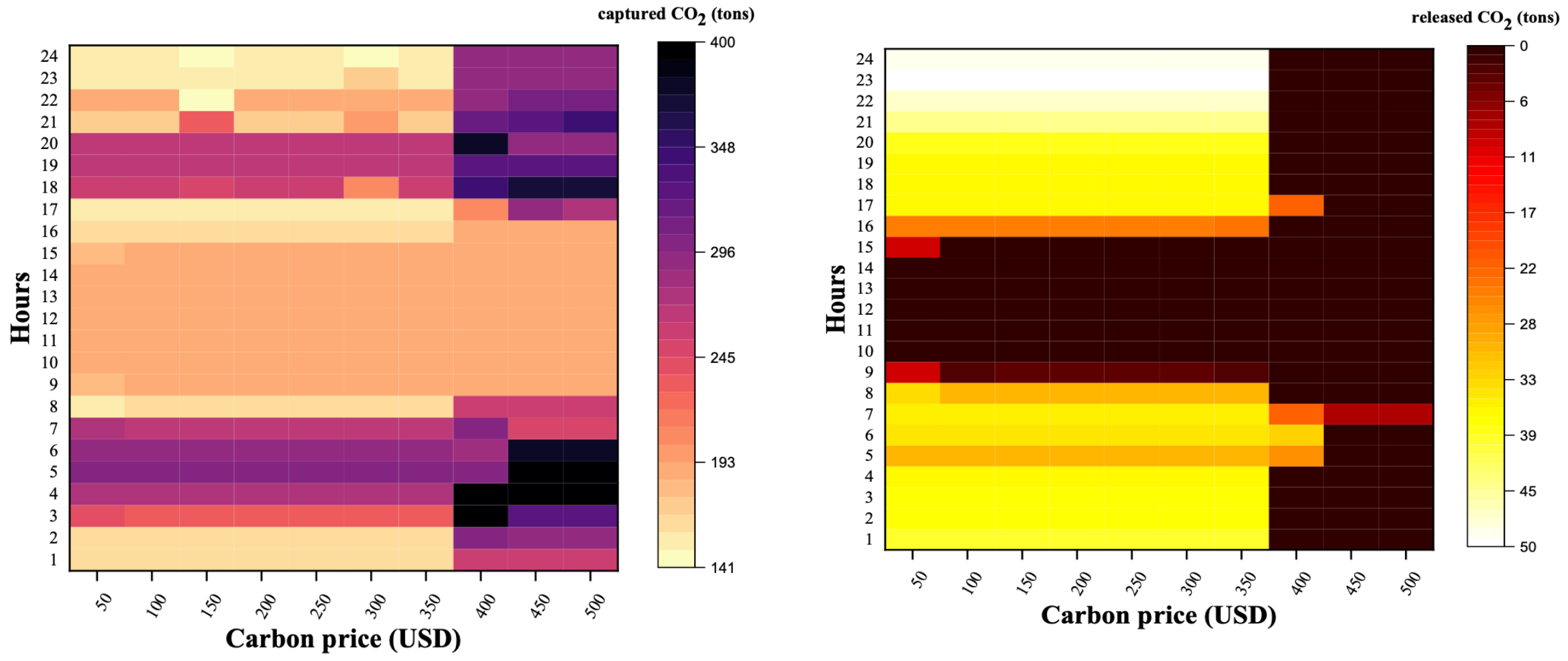

**Figure 16.** Carbon price influence on CDRT.

*4.9. Proposed* DRL-agent comparison

The proposed DRL agent in this study is compared to other agents and a rule-based optimisation. The particle swarm optimisation (PSO)-based stochastic optimisation method, a classical heuristic algorithm, is set as the rule-based benchmark. It is worth stating that the mechanism of the stochastic PSO is to solve the stochastic problem by taking the average of the optimal actions associated with a given number of samples at each time slot. Thus, the action sample is set to 200, while the particles and maximum iterations are set to 30 and 100 [36], respectively. Emphatically, the PSO objective function is the systems' overall economic cost in eq (30) with the energy balance and ramp limit constraints in eq (33) – eq (36), including the specification of the variables (i.e., each systems' output) boundary limit. For the SAC and TD3 agents without hyperparameter tuning features, the default parameters by stable-baseline3 [40] are adopted. The proposed systems' configuration (i.e., ZCMES with PCCS and solvent-based DAC) is the environment for evaluating the methods. Table 3 shows the comparative results of the methods. It can be observed that the RL-based methods, i.e., SAC and TD3 achieved better performance compared to the PSO-based stochastic optimisation method in minimising economic cost. The reason is that PSO easily falls into a local optimum in high dimensional space, which is the situation with our ZCMES environment [50]. Another notable observation is that the RL-base methods with automated features outperformed others which validates the influence of hyperparameters tuning. Meanwhile, our proposed approach achieved the best performance with a 20.89% improvement compared to PSO. This is achieved by combining its entropy feature to navigate through high-dimensional space and utilising suitable hyperparameter values by the auto feature that enables fast convergence and an optimum global prospect.

**Table 3.** Different approaches comparison results

| Method | Operation cost ($) | Fuel cost ($) | Emission cost ($) | CDR cost ($) | Overall cost ($) | Improvement |
|---|---|---|---|---|---|---|
| PSO | 95,085.58 | 153,296.48 | 28,108.33 | 830,389.37 | 1,105,540.28 | - |
| SAC | 77,110.10 | 124,316.51 | 22,794.57 | 673,408.22 | 896,543.17 | 18.90% |
| TD3 | 79,177.76 | 127,649.98 | 23,405.79 | 691,465.22 | 920,583.39 | 16.73% |
| a-TD3 | 75,222.20 | 121,272.85 | 22,236.49 | 654,921.03 | 874,592.92 | 20.89% |
| a-SAC* | 72,597.88 | 117,041.93 | 21,460.71 | 634,002.64 | 844,080.48 | **23.65%** |

*a: auto; *proposed method*

## 5.0 Conclusion

This study proposes a novel data-driven control strategy for the real-time scheduling of IES with carbon capture technology. The control strategy is based on DRL with automated hyperparameter tuning that is implemented with Optuna. SAC and TD3 are compared using the proposed framework as they are the most adopted model-free DRL approach for continuous action spaces. The results show that the SAC algorithm outperformed TD3, even with the hyperparameter tuning, and the proposed auto-SAC DRL agent outperformed the rule-based method with 23.65% improvement.

To evaluate the influence of CDR technology as a carbon elimination potential, PCC and DAC technology are integrated with MES, and various possible configurations were evaluated and compared. Notably, real multi-energy data from a district community in Arizona, United States, is selected as a case study. In addition, the economic, technical, and chemical properties of the considered CDRT were obtained from the renowned manufacturers' catalogue and pilot study project documentation. Furthermore, we proposed two important CDRT metrics (CRI and CCRR), where it is expected that a superior configuration will have a low CRI value and a high CCRR value. Finally, we applied the hyperparameter values selected by the Optuna to train DRL agents for each considered scenario in an offline mode. The trained agents were tested and deployed online. The obtained simulation results show that the agents can meet the multi-energy demand of the prosumers without any physical constraint violation. Various scenario analyses were also presented to compare the proposed CDRT configuration. The obtained result affirmed that PCC and solvent-based DAC coupling is the most suitable option with the least CDR cost of 31.54% reduction, a high CCRR value of 38.54 that indicates a high capturing ratio, and a low carbon release indicator (CRI) of 2.53. The reason is due to the low-temperature thermal demand that the waste heat of MES can easily provide and the high absorption capacity of the chosen sorbent compared to an aqueous solvent. However, CDRT is not economically viable at the current carbon price; we conducted a sensitivity analysis that shows CDRT will be financially feasible at a carbon price of 400-450USD/ton considering its present economic, technical, and chemical properties. However, the provision of incentives to encourage the adoption of CDRT in the energy sector cannot be excluded.

**Appendix A**

***Renewable energy model***

The mathematical model governing WT and PV energy generation output is described in Eq.A1 – Eq.A4. Eq.A1 is the power output of PV, while Eq. A2 is the power output of the WT and Eq. A3 is the sum of the total renewable energy generation.

$$P_{PV,t} = \eta_v . A_{pv} . G_t\left[1 + 0.001[T_C - T_{STC}]\right] \tag{A1}$$

$$P_{wt} = N_{wt}\begin{cases} \frac{P_{wt}^{-}(v - v_{ci})}{\bar{v} - v_{ci}} & v_{ci} \leq v \leq \bar{v} \\ P_{wt}^{-} & \bar{v} \leq v \leq v_{co} \\ 0 & else \end{cases} \tag{A2}$$

$$\boldsymbol{P_{RES,t}} = P_{PV,t} + P_{wt} \tag{A3}$$

$$\boldsymbol{P_{RES,t}} = P_{RES\rightarrow EL,t} + P_{RES\rightarrow EC,t} + P_{RES\rightarrow WSHP,t} + P_{RES\rightarrow BES,t} + P_{RES\rightarrow DAC,t} + P_{RES\rightarrow PCCS,t} \tag{A4}$$

***Cogeneration plant***

This study considers gas-fired and coal-fired power plants (CFP) cogeneration plants for producing electricity and thermal energy. Eq. A6 – Eq. A10 describes the gas-fired plant model considering its energy output (electricity and thermal), the PLR consideration described in the manuscript, ramp limit constraint, the energy distribution within MES, and carbon emission estimation.

$$P_{GT(t)}^{e} = P_{CHP}^{e,r}\varphi_{CHP(t)}^{e}/\eta_{gt}^{e,r} \tag{A5}$$

$$\varphi_{CHP(t)}^{e} = \delta_{CHP(t)}\Big/\sum_{o\epsilon O_{CHP}^{e}} k_{CHP,i}^{e}\delta_{CHP(t)}^{i} \tag{A6}$$

$$u_{gt(t)}\delta_{GT,min} \leq \delta_{GT(t)} \leq u_{gt(t)}\delta_{GT,max} \tag{A7}$$

$$\Pi_{gas,t}^{CHP} = \left(\frac{P_{CHP}^{e,r}\delta_{GT(t)}}{\eta_{gt}^{e,r}}\right)/LHV \tag{A8}$$

$$P^{h}_{CHP(t)} = P^{e,r}_{CHP}\eta^{h,r}_{gt}\varphi^{e}_{CHP(t)}/\eta^{e,r}_{gt} \tag{A9}$$

$$\varphi^{h}_{CHP(t)} = \delta_{CHP(t)}\sum_{o\epsilon O^{h}_{CHP}} k^{h}_{CHP,o}\delta^{o}_{CHP(t)} \Big/ \sum_{o\epsilon O^{e}_{CHP}} k^{e}_{CHP,o}\delta^{o}_{CHP(t)} \tag{A10}$$

$$-R \le P_{GT(t-1)} - P_{GT(t)} \le R \tag{A11}$$

$$P^{e}_{CHP(t)} = P^{e}_{CHP\to EL,t} + P^{e}_{CHP\to EC,t} + P^{e}_{CHP\to WSHP,t} + P^{e}_{CHP\to BES,t} + P^{e}_{CHP\to PCCS,t} + P^{e}_{CHP\to DAC,t} \tag{A12}$$

$$P^{h}_{CHP(t)} = P^{h}_{CHP\to QL,t} + P^{h}_{CHP\to TES,t} + P^{h}_{CHP\to AC,t} + P^{h}_{CHP\to DAC,t} \tag{A13}$$

$$GT_{CO2(t)} = P^{e}_{GT(t)}.F_{CO2,gt} \tag{A14}$$

***Coal-fired power plant model***

A coal-fired plant (CFP) is another cogeneration plant that consumes coal for power and thermal generation production. The model is described in Eq. A15 – Eq. A19 and the carbon emission during operation is computed in Eq. A20.

$$P^{CHP}_{e,t} = \delta_{CHP(t)}.P^{r}_{CHP} \tag{A15}$$

$$u_{CHP(t)}\delta_{CHP,min} \le \delta_{CHP(t)} \le u_{CHP(t)}\delta_{CHP,max} \tag{A16}$$

$$P_{coal,t} = P^{CHP}_{e,t} \Big/ \eta_{CHP,e}HV_{coal} \tag{A17}$$

$$H^{CHP}_{q,t} = \eta_{CHP,q}P_{coal,t}HV_{coal} \tag{A18}$$

$$-R \le P_{CHP(t-1)} - P_{CHP(t)} \le R \tag{A19}$$

$$D^{CfP}_{CO2(t)} = P_{coal,t}.F_{CO2,cfp} \tag{A20}$$

***Gas boiler and water-source heat pump model***

In this study, the heat demand by the prosumers is met by GB and WSHP. The thermal output of the two devices is calculated in Eq. A21, while the PLR function is computed in Eq. A22 and Eq. A23 expressed the loading rate boundary. As popularly known, GB consumes natural gas (NG) to produce thermal energy; the NG consumption at each time step is computed in Eq. A24, while the associated $CO_2$ for burning NG is estimated in Eq. A25. Further, the power input of WSHP and the optimal input from associated devices is described in Eq. A26 and Eq. A27, respectively. Finally, the thermal output distribution of GB and WSHP is illustrated in Eq. A28 and Eq. A29, respectively.

$$Q^{h}_{GB/WSHP(t)} = Q^{r}_{GB/WSHP}\varphi^{h}_{GB/WSHP(t)}/COP^{r}_{GB/WSHP} \tag{A21}$$

$$\varphi^{h}_{GB/WSHP(t)} = \delta_{GB/WSHP(t)} \Big/ \sum_{o\epsilon O_{GB/WSHP}} k_{GB/WSHP,o}\delta^{o}_{GB/WSHP(t)} \tag{A22}$$

$$u_{gb/WSHP(t)}\delta_{GB/WSHP,min} \leq \delta_{GB/WSHP(t)} \leq u_{gb/WSHP(t)}\delta_{GB/WSHP,max} \tag{A23}$$

$$\Pi^{GB}_{gas,t} = Q^{r}_{GB}\delta_{GB(t)}/COP^{r}_{GB}/LHV \tag{A24}$$

$$GB_{CO2(t)} = \Pi^{GB}_{gas,t} * co_2 \tag{A25}$$

$$-R \leq Q^{h}_{GB/WSHP(t-1)} - Q^{h}_{GB/WSHP(t)} \leq R \tag{A26}$$

$$P^{in}_{WSHP(t)} = Q^{r}_{HB}\delta_{GB(t)}/COP^{r}_{HB} \tag{A27}$$

$$P^{in}_{WSHP(t)} = P_{RES\to WSHP,t} + P_{BES\to WSHP,t} + P^{e}_{CHP\to WSHP,t} \tag{A28}$$

$$Q^{h}_{GB(t)} = P_{GB\to QL,t} + P_{GB\to AC,t} + P_{GB\to TES,t} + P_{GB\to DAC,t} \tag{A29}$$

$$Q^{h}_{WSHP(t)} = P_{WSHP\to QL,t} + P_{WSHP\to AC,t} + P_{WSHP\to TES,t} + P_{WSHP\to DAC,t} \tag{A30}$$

***Electric chiller and absorption chiller model***

EC and AC are introduced to meet the cooling demand of the prosumers. Eq. A31 estimates the cooling output of the two (2) equipment where $Q^{r}_{AC/EC}$ is the rated capacity, considering PLR influence on the on-design condition as described in section 2.3 and computed in Eq. A32. The loading rates of the equipment are described in Eq. A33 where $u_{AC/EC(t)}$ is the binary variable for controlling the equipment status. The power input into the system for the specific cooling production is computed in Eq. A34. Basically, EC is supplied with electric power (from RES, CHP or BES) for the cooling generation while EC consumes thermal energy, which can be supplied by waste heat from CHP and the thermal output of WSHP, GB and TES. Thus, the optimal power input combination is described in Eq. A36 and Eq. A37.

$$Q^{c}_{AC/EC(t)} = Q^{r}_{AC/EC}\varphi^{c}_{AC/EC(t)}/COP^{r}_{AC/EC} \tag{A31}$$

$$\varphi^{c}_{AC/EC(t)} = \delta_{AC/EC(t)} \Big/ \sum_{o\epsilon O_{AC/EC}} k_{AC/EC,o}\delta^{o}_{AC/EC(t)} \tag{A32}$$

$$u_{AC/EC(t)}\delta_{AC,min} \leq \delta_{AC/EC(t)} \leq u_{AC/EC(t)}\delta_{AC,max} \tag{A33}$$

$$-R \leq Q^{c}_{AC/EC(t-1)} - Q^{c}_{AC/EC(t)} \leq R \tag{A34}$$

$$P^{in}_{AC/EC(t)} = Q^{r}_{AC/EC}\delta_{AC/EC(t)}/COP^{r}_{AC/EC} \tag{A35}$$

$$P^{in}_{AC(t)} = P^{h}_{CHP\to AC,t} + P_{GB\to AC,t} + P_{WSHP\to AC,t} + P_{TES\to AC,t} \tag{A36}$$

$$P^{in}_{EC(t)} = P_{RES\to EC,t} + P^{e}_{CHP\to EC,t} + P_{BES\to AC,t} \tag{A37}$$

**Table A1.** Installed equipment capacity and associated operation cost

| | Installed capacity (MW) | Operation cost ($/MWh) |
|---|---|---|
| PV | - | 6.92 |
| WT | 100nos | 6.92 |
| GT | 500 | 2.70 |
| CFP | 500 | 4.80 |
| WSHP | 500 | 1.70 |
| EC | 200 | 1.50 |
| AC | 200 | 1.50 |
| PCC | 300tons | - |
| DAC | 200tons | - |
| BES | 100 | 1.00 |
| TES | 200 | 1.00 |

**PV**: photovoltaic system; **WT**: wind turbine; **GT**: gas turbine; **CFP**: coal-fired plant; **WSHP**: water-source heat pump; **EC**: electric chiller; **AC**: absorption chiller; **PCC**: Post carbon capture system; **DAC**: Direct-air capture system; **BES**: Battery electrical system; **TES**: Thermal energy storage.

***Acronyms***

RES: Renewable energy

Pec: Electric chiller power input

WSHP: water-source heat pump

Php: WSHP power input

Gte: GT electric power output

Pcoale: coal-fired plant electric power output

Pbch: charging power of BES

Pbdch: discharging power of BES

Ptch: charging power of TES

Ptdch: discharging power of TES

PCCel: electric demand of PCC

DACel: electric demand of DAC

DACh: thermal demand of DAC

Qhp: thermal output of WSHP

Qgb: thermal output of gas boiler

Pcoalq: thermal output of coal-fired plant

## 6.0 References

[1] T. M. Alabi *et al.*, "A review on the integrated optimization techniques and machine learning approaches for modeling, prediction, and decision making on integrated energy systems," *Renewable Energy,* vol. 194, no. C, 2022, doi: 10.1016/j.renene.2022.05.123.

[2] T. M. Alabi, L. Lu, Z. Yang, and Y. Zhou, "A novel optimal configuration model for a zero-carbon multi-energy system (ZC-MES) integrated with financial constraints," *Sustainable Energy, Grids and Networks,* vol. 23, 2020, doi: 10.1016/j.segan.2020.100381.

[3] T. M. Alabi, L. Lu, and Z. Yang, "Stochastic optimal planning scheme of a zero-carbon multi-energy system (ZC-MES) considering the uncertainties of individual energy demand and renewable resources: An integrated chance-constrained and decomposition algorithm (CC-DA) approach," *Energy,* vol. 232, 2021, doi: 10.1016/j.energy.2021.121000.

[4] T. M. Alabi, F. D. Agbajor, Z. Yang, L. Lu, and A. J. Ogungbile, "Strategic potential of multi-energy system towards carbon neutrality: A forward-looking overview," *Energy and Built Environment,* 2022/06/20/ 2022, doi: https://doi.org/10.1016/j.enbenv.2022.06.007.

[5] IRENA. "Renewable Capacity Statistics 2020." International Renewable Energy Agency. https://irena.org/publications/2020/Mar/Renewable-Capacity-Statistics-2020 (accessed October 11, 2021).

[6] V. Sebestyén, "Renewable and Sustainable Energy Reviews: Environmental impact networks of renewable energy power plants," *Renewable and Sustainable Energy Reviews,* vol. 151, 2021, doi: 10.1016/j.rser.2021.111626.

[7] J. W. Busby *et al.*, "Cascading risks: Understanding the 2021 winter blackout in Texas," *Energy Research & Social Science,* vol. 77, 2021, doi: 10.1016/j.erss.2021.102106.

[8] M. J. Burke and J. C. Stephens, "Political power and renewable energy futures: A critical review," *Energy Research & Social Science,* vol. 35, pp. 78-93, 2018, doi: 10.1016/j.erss.2017.10.018.

[9] T. M. Alabi, L. Lu, and Z. Yang, "A novel multi-objective stochastic risk co-optimization model of a zero-carbon multi-energy system (ZCMES) incorporating energy storage aging model and integrated demand response," *Energy,* vol. 226, no. C, 2021, doi: 10.1016/j.energy.2021.120258.

[10] T. M. Alabi, L. Lu, and Z. Yang, "Data-driven optimal scheduling of multi-energy system virtual power plant (MEVPP) incorporating carbon capture system (CCS),

electric vehicle flexibility, and clean energy marketer (CEM) strategy," *Applied Energy,* vol. 314, 2022, doi: 10.1016/j.apenergy.2022.118997.

[11] J. Liu, S. Cao, X. Chen, H. Yang, and J. Peng, "Energy planning of renewable applications in high-rise residential buildings integrating battery and hydrogen vehicle storage," *Applied Energy,* vol. 281, 2021, doi: 10.1016/j.apenergy.2020.116038.

[12] N. MacDowell *et al.*, "An overview of CO2 capture technologies," *Energy & Environmental Science,* vol. 3, no. 11, 2010, doi: 10.1039/c004106h.

[13] U. D. o. Energy, *Carbon-Dioxide-Capture Handbook*, Mongarton, WV: National Energy Technology Laboratory 2015. [Online]. Available: www.netl.doe.gov.

[14] F. Font, T. G. Myers, and M. G. Hennessy, "A Mathematical Model of Carbon Capture by Adsorption," in *Multidisciplinary Mathematical Modelling*, (SEMA SIMAI Springer Series, 2021, ch. Chapter 3, pp. 35-48.

[15] C. Dhoke *et al.*, "Sorbents screening for post-combustion CO2 capture via combined temperature and pressure swing adsorption," *Chemical Engineering Journal,* vol. 380, 2020, doi: 10.1016/j.cej.2019.122201.

[16] J. Yang, N. Zhang, Y. Cheng, C. Kang, and Q. Xia, "Modeling the Operation Mechanism of Combined P2G and Gas-Fired Plant With CO2 Recycling," *IEEE Transactions on Smart Grid,* vol. 10, no. 1, pp. 1111-1121, 2019, doi: 10.1109/tsg.2018.2849619.

[17] G. Zhang *et al.*, "Modeling and Optimization of Integrated Energy System for Renewable Power Penetration considering Carbon and Pollutant Reduction Systems," *Frontiers in Energy Research,* vol. 9, 2021, doi: 10.3389/fenrg.2021.767277.

[18] X. Liu, X. Li, J. Tian, and H. Cao, "Low-carbon economic dispatch of integrated electricity and natural gas energy system considering carbon capture device," *Transactions of the Institute of Measurement and Control,* 2021, doi: 10.1177/01423312211060572.

[19] X. Zhang, Y. Bai, and Y. Zhang, "Collaborative optimization for a multi-energy system considering carbon capture system and power to gas technology," *Sustainable Energy Technologies and Assessments,* vol. 49, 2022, doi: 10.1016/j.seta.2021.101765.

[20] G. Zhang, W. Wang, Z. Chen, R. Li, and Y. Niu, "Modeling and optimal dispatch of a carbon-cycle integrated energy system for low-carbon and economic operation," *Energy,* vol. 240, 2022, doi: 10.1016/j.energy.2021.122795.

[21] M. Berger, D. Radu, R. Fonteneau, T. Deschuyteneer, G. Detienne, and D. Ernst, "The role of power-to-gas and carbon capture technologies in cross-sector decarbonisation strategies," *Electric Power Systems Research,* vol. 180, 2020, doi: 10.1016/j.epsr.2019.106039.

[22] G. Realmonte *et al.*, "An inter-model assessment of the role of direct air capture in deep mitigation pathways," *Nat Commun,* vol. 10, no. 1, p. 3277, Jul 22 2019, doi: 10.1038/s41467-019-10842-5.

[23] T. Terlouw, K. Treyer, C. Bauer, and M. Mazzotti, "Life Cycle Assessment of Direct Air Carbon Capture and Storage with Low-Carbon Energy Sources," *Environmental Science & Technology,* vol. 55, no. 16, pp. 11397-11411, 2021, doi: 10.1021/acs.est.1c03263.

[24] M. M. J. de Jonge, J. Daemen, J. M. Loriaux, Z. J. N. Steinmann, and M. A. J. Huijbregts, "Life cycle carbon efficiency of Direct Air Capture systems with strong hydroxide sorbents," *International Journal of Greenhouse Gas Control,* vol. 80, pp. 25-31, 2019, doi: 10.1016/j.ijggc.2018.11.011.

[25] C. Engineering. https://carbonengineering.com/ (accessed March 5, 2022).

[26] Climeworks. https://climeworks.com/ (accessed April 4, 2022).

[27] Climeworks. "Climeworks launches direct air capture plant in Italy." Climeworks. https://climeworks.com/news/climeworks-launches-dac-3-plant-in-italy (accessed April 10, 2022).
[28] M. T. Mota-Martinez, J. P. Hallett, and N. Mac Dowell, "Solvent selection and design for CO2capture – how we might have been missing the point," *Sustainable Energy & Fuels,* vol. 1, no. 10, pp. 2078-2090, 2017, doi: 10.1039/c7se00404d.
[29] S. Deutz and A. Bardow, "Life-cycle assessment of an industrial direct air capture process based on temperature–vacuum swing adsorption," *Nature Energy,* vol. 6, no. 2, pp. 203-213, 2021, doi: 10.1038/s41560-020-00771-9.
[30] T. M. Alabi, L. Lu, and Z. Yang, "Improved hybrid inexact optimal scheduling of virtual powerplant (VPP) for zero-carbon multi-energy system (ZCMES) incorporating Electric Vehicle (EV) multi-flexible approach," *Journal of Cleaner Production,* 2021, doi: 10.1016/j.jclepro.2021.129294.
[31] T. Yang, L. Zhao, W. Li, and A. Y. Zomaya, "Reinforcement learning in sustainable energy and electric systems: a survey," *Annual Reviews in Control,* Review vol. 49, pp. 145-163, 2020, doi: 10.1016/j.arcontrol.2020.03.001.
[32] A. H. Ganesh and B. Xu, "A review of reinforcement learning based energy management systems for electrified powertrains: Progress, challenge, and potential solution," *Renewable and Sustainable Energy Reviews,* vol. 154, 2022, doi: 10.1016/j.rser.2021.111833.
[33] Y. Huang, P. Ding, Y. Wang, S. Li, K. Yang, and Y. Li, "A bilevel optimal operation model of multi energy carriers system considering part load rate and demand response," *Sustainable Energy Technologies and Assessments,* vol. 45, 2021, doi: 10.1016/j.seta.2021.101035.
[34] Q. H. Wu *et al.*, "Comparison and error analysis of off-design and design models of energy hubs," *CSEE Journal of Power and Energy Systems,* 2019, doi: 10.17775/cseejpes.2018.00630.
[35] T. P. Lillicrap *et al.*, "Continuous control with deep reinforcement learning," *CoRR,* vol. abs/1509.02971, 2016.
[36] B. Zhang *et al.*, "Soft actor-critic –based multi-objective optimized energy conversion and management strategy for integrated energy systems with renewable energy," *Energy Conversion and Management,* Article vol. 243, 2021, Art no. 114381, doi: 10.1016/j.enconman.2021.114381.
[37] B. Zhang, W. Hu, D. Cao, Q. Huang, and Z. Chen, "Asynchronous Advantage Actor-Critic Based Approach for Economic Optimization in the Integrated Energy System with Energy Hub," in *2021 3rd Asia Energy and Electrical Engineering Symposium, AEEES 2021*, 2021, pp. 1170-1176, doi: 10.1109/AEEES51875.2021.9403176. [Online]. Available: https://www.scopus.com/inward/record.uri?eid=2-s2.0-85105293022&doi=10.1109%2fAEEES51875.2021.9403176&partnerID=40&md5=d8dceaf5e880d8c63b2b624267042a2b
[38] T. Haarnoja, A. Zhou, P. Abbeel, and S. Levine, "Soft Actor-Critic: Off-Policy Maximum Entropy Deep Reinforcement Learning with a Stochastic Actor," presented at the Proceedings of the 35th International Conference on Machine Learning, Proceedings of Machine Learning Research, 2018. [Online]. Available: https://proceedings.mlr.press/v80/haarnoja18b.html.
[39] T. Haarnoja, H. Tang, P. Abbeel, and S. Levine, "Reinforcement Learning with Deep Energy-Based Policies," presented at the Proceedings of the 34th International Conference on Machine Learning, Proceedings of Machine Learning Research, 2017. [Online]. Available: https://proceedings.mlr.press/v70/haarnoja17a.html.

[40] S. D. C. Team. *Metadata record for: The Building Data Genome Project 2, energy meter data from the ASHRAE Great Energy Predictor III competition*, 26-Oct-2020, doi: doi: 10.6084/m9.figshare.13033847.v1. .
[41] C. A. M. Services. "CAMS solar radiation time-series." https://ads.atmosphere.copernicus.eu/cdsapp#!/dataset/cams-solar-radiation-timeseries?tab=form (accessed July 20, 2021).
[42] "Affordable Clean Energy (ACE) Rule." United States Environmental Protection Agency. https://www.epa.gov/stationary-sources-air-pollution/proposal-affordable-clean-energy-ace-rule (accessed Jan 15, 2022).
[43] D. W. Keith, G. Holmes, D. St. Angelo, and K. Heidel, "A Process for Capturing CO2 from the Atmosphere," *Joule,* vol. 2, no. 8, pp. 1573-1594, 2018, doi: 10.1016/j.joule.2018.05.006.
[44] "Petra Nova - W.A. Parish Project." Office of Fossil Energy and Carbon Management. https://www.energy.gov/fecm/petra-nova-wa-parish-project#:~:text=The%20Petra%20Nova%20CCS%20project,of%20this%20greenhouse%20gas%20annually. (accessed 2022, Jan 14).
[45] J. I. Huertas, M. D. Gomez, N. Giraldo, and J. Garzón, "CO2Absorbing Capacity of MEA," *Journal of Chemistry,* vol. 2015, pp. 1-7, 2015, doi: 10.1155/2015/965015.
[46] W. J. Schmelz, G. Hochman, and K. G. Miller, "Total cost of carbon capture and storage implemented at a regional scale: northeastern and midwestern United States," *Interface Focus,* vol. 10, no. 5, p. 20190065, Oct 6 2020, doi: 10.1098/rsfs.2019.0065.
[47] X. Shi *et al.*, "Sorbents for the Direct Capture of CO2 from Ambient Air," *Angew Chem Int Ed Engl,* vol. 59, no. 18, pp. 6984-7006, Apr 27 2020, doi: 10.1002/anie.201906756.
[48] M. Zhan, J. Chen, C. Du, and Y. Duan, "Twin Delayed Multi-Agent Deep Deterministic Policy Gradient," presented at the 2021 IEEE International Conference on Progress in Informatics and Computing (PIC), 2021.
[49] "Optuna - A hyperparameter optimization framework." Optuna. https://optuna.org/ (accessed Jan 10, 2022).
[50] M. Li, W. Du, and F. Nian, "An Adaptive Particle Swarm Optimization Algorithm Based on Directed Weighted Complex Network," *Mathematical Problems in Engineering,* vol. 2014, pp. 1-7, 2014, doi: 10.1155/2014/434972.